\documentclass[]{jfm}

\usepackage{siunitx}
\usepackage{graphicx}
\usepackage{newtxtext}
\usepackage{newtxmath}
\usepackage{natbib}
\usepackage[normalem]{ulem}
\usepackage{hyperref}
\hypersetup{
    colorlinks = true,
    urlcolor   = blue,
    citecolor  = black,
}

\newcommand{\RomanNumeralCaps}[1]
\linenumbers
\usepackage{bm}
\usepackage{multirow}
\usepackage{amsmath,amssymb}
\usepackage[colorinlistoftodos]{todonotes}

\newcommand{\am}{\textcolor{black}}

\title{Effects of permeability on hindered settling of porous particles
}
\author{Alexander Metelkin\aff{1}
  \and Bernhard Vowinckel\aff{1}\corresp{\email{bernhard.vowinckel@tu-dresden.de}}}

\affiliation{\aff{1}Institute of Urban and Industrial Water Management, Technische Universität Dresden, Bergstraße 66, Dresden, 01069, Germany}

\begin{document}
\maketitle
\begin{abstract}
We investigate the settling behavior of suspensions of highly porous and permeable particles in the viscous regime using particle-resolved direct numerical simulations (DNS). The simulations employ a coupled Euler-Lagrange framework that accounts for particle permeability. The results show that the settling behavior of permeable particles follows the classical power-law relationship of Richardson–Zaki in terms of their settling velocity, but particles with higher permeability settle faster as the particle volume fraction increases. At a particle volume fraction of 30\%, the difference in settling speed is up 106\% between the least and most permeable particles investigated in this study. We explain this effect by the alteration of counter flows induced by the fluid displacement of settling particles.
Quantitative analysis of the mean vertical fluid velocity confirms that suspensions composed of more permeable particles generate weaker counterflows, posing less resistance to the settling motion.
We furthermore show how velocity fluctuations and self-diffusivity depend on the permeability of the porous particles and the particle volume fraction. Both quantities increase with volume fraction and are largest for the least permeable particles, except at the highest volume fraction where reduced settling velocities reverse this trend.
The influence of particle permeability also reveals two effects in the particle microstructure. First, the analysis showed that particle clustering decreases with increasing permeability. Second, the overall probability of finding a neighbor within the lubrication range is lowest for the least permeable particles We attribute this to the weakening of repulsive pressure forces between tumbling particles with increased permeability.%Stop
\end{abstract}

%\begin{keywords}
%\am{DO NOT KEEP THEM HERE FOR SUBMISSION!
%Permeable particles dynamics, hindered settling, suspension mechanics, particle-resolved direct numerical simulation}
%\am{Authors should not enter keywords on the manuscript, as these must be chosen by the author during the online submission process and will then be added during the typesetting process (see \href{https://www.cambridge.org/core/journals/journal-of-fluid-mechanics/information/list-of-keywords}{Keyword PDF} for the full list).  Other classifications will be added at the same time.}

%\end{keywords}

\section{Introduction}
\label{sec:introduction}

The settling behavior of fine-grained sediments (i.e., grains with diameters smaller than $63 \mu m$) is central to hydrological and environmental processes, influencing the morphology of estuaries, deltas, rivers, and coasts \citep{dyer1989sediment,seminara2001river,mcanally2001coastal}. 
Owing to their small size, fine-grained particles are cohesive and assemble into larger porous structures known as flocs, thereby altering their settling behavior \citep{manning2004development,vercruysse2017suspended,vowinckel2023investigating}.
The interaction of these porous flocs with their ambient fluid has multifaceted consequences for environmental processes such as the marine carbon cycle, considering that sediments are a source of organic matter and minerals \citep{alldredge1988characteristics}, as well as a carrier of contaminants \citep{droppo2001rethinking,grabowski2011erodibility,gillard2019}. 
An accurate representation of this interaction requires detailed characterization of floc properties, including density, shape, porosity, and permeability, that govern the drag and determine the settling speed and transport behavior \citep{french2010critical,strom2011explicit,spearman2020measurement}.
The suspension settling behavior is also a key mechanism in the area of wastewater treatment. In treatment plants, sludge flocs circulate in suspension before being separated in clarifiers; their settling behavior in both dilute (where sediment volume fraction $\phi \lesssim 10^{-2}$) and dense ($\phi \gtrsim 10^{-2} $) regimes determines how effectively water can be clarified \citep{JARVIS20053121,de2008extending,franccois2016experimental,CUI2019159,roche2022densification}.

While the settling behavior of non-cohesive coarse sediments is fairly well established, the settling dynamics of fine-grained sediments such as silt and clay is more complex \citep{grabowski2011erodibility}. Such particles are subject to cohesive van der Waals forces that dominate over gravitational forces if particles are close to each other \citep{vowinckel2019settling}. 
%\bv{BV: I do not understand "gravitational effects at the length scale comparable with their size". What do you mean?}
Under the influence of such forces, clay platelets flocculate into aggregates known as flocculi, which cannot be broken down further under local shear conditions and thus form the primary particles of a given size distribution \citep{winterwerp2004introduction}. Subsequently, these primary particles can combine into larger structures, forming flocs, which may further combine into even larger floc aggregates. This process follows a four-level aggregation hierarchy: clay platelets, flocculi, flocs, and floc aggregates \citep{leussen1988aggregation}. The hierarchical structure of flocculation forms porous aggregates with different porosities and permeabilities, posing challenges in predicting their settling velocity.

%\sout{the terms  can be collectively denoted as porous particles for the sake of generalization.} 
Observations indicate that the majority of fine-grained sediments is transported in the flocculated state as porous aggregates
%\sout{sediment mass within cohesive suspensions is contained in aggregates}
\citep{kranck1992characteristics,lamb2020mud,krahl2022impact}, which settle at speeds that exceed those of their constituent primary particles \citep{manning2006variability}. Owing to their complex internal porous composition, the density of aggregates is difficult to determine directly. Their small size and fragile structure hinder extraction without structural damage, rendering invasive experimental methods impractical \citep{eisma1997situ,manning2006variability}. Consequently, physical attributes such as floc density are derived via settling velocity models, which are usually expressed as algebraic relations that determine the settling rate based on sediment properties \citep{manning2006variability,strom2011explicit} but still bear a high level of uncertainty. In the scope of this study, we deal with the settling of porous particles and generalize flocs and floc aggregates as individual porous particles.
%\sout{which characterize the dependence of settling rate on sediment properties.}
%\bv{BV: what is a linear settling velocity model? You should provide a few citations here.}
%\am{Agree that term linear could be misleading, changed it to "algebraic relation".}

%However, this approach neglects the influence of permeability, which can significantly alter the effective drag on highly porous aggregates \citep{metelkin2025parameters}.

%\bv{You should be more systematic in your literature research starting with settling velocity models for single particles but then you should quickly bridge the gap to more dense systems, because this is the main focus of your study}
%\idea{Introducing settling velocity models for the dilute regime}

There are various settling velocity models developed for coarse and fine grained sediments in the dilute regime. The simplest approaches invoke a force balance between the submerged weight of a spherical particle and the Stokes drag. These models are valid only in the creeping-flow regime, i.e. under conditions for which the Reynolds number based on the single-particle settling velocity satisfies $\Rey_{\parallel,0} = U_{\parallel,0} D / \nu \ll 1$, where $ U_{\parallel,0}$ denotes the converged settling velocity of an isolated sediment particle, $D$ its diameter, and  $\nu$ the kinematic viscosity of the ambient fluid. To extend the applicability of settling velocity models beyond the creeping-flow regime, \citet{ferguson2004simple} introduced a unified expression that combines laminar and turbulent drag contributions. Subsequently, for fine-grained sediment aggregates, \citet{strom2011explicit} proposed an augmented model that accounts for the decrease in floc density with increasing size of the floc and incorporates empirical coefficients accounting for aggregate shape and permeability.

For the collective settling of particles, models for the dense regime were developed by explicitly accounting for the particle volume fraction within the suspension. Settling in dense suspensions is considerably more complex than in the dilute regime owing to several additional mechanisms. These include counterflows generated by particles displacing fluid, wake-mediated hydrodynamic interactions that influence the motion of neighbouring particles, direct particle–particle collisions, and cluster formation \citep{winterwerp2004introduction}. Owing to these additional mechanisms, which reduce the bulk settling velocity of the sediment, settling in dense suspensions is commonly referred to as hindered settling.
The hindered settling of monodisperse solid spheres in the dense regime was examined in the seminal work by \citet{richardson1954sedimentation}. The authors proposed a semi-empirical formulation (denoted here as the RZ equation), in which the settling velocity follows a power law dependence on the suspension voidage, defined as $(1-\phi)$, with a constant exponent. Later, \citet{richardson1954sedfluid} showed that this exponent is a function of the Reynolds number $\Rey_{\parallel,0}$, and subsequent experimental studies \citep{garside1977velocity,BALDOCK200491} further refined correlations for the exponent as a function of $\Rey_{\parallel,0}$.
%\bv{You essentially mean $Re_{\parallel,0}$, right?} \am{Yes, should I mention this explicitly?}
The applicability of the RZ equation was also confirmed by many numerical studies of spherical particles settling in a viscous fluid across a wide range of Reynolds numbers $Re_{\parallel,0}$ and volume fractions \citep{yin2007hindered,hamid2014direct,yao2021effects}. However, at low particle volume fractions, numerical results \citep{yin2007hindered,shajahan2020influence} revealed clear deviations from the RZ equation, indicating that for $\phi \lesssim 0.05$ the settling velocity exhibits a steeper scaling than in denser suspensions with $0.3>\phi > 0.05$. For cohesive suspensions, \citet{mehta1986characterization} adapted the RZ equation, and \citet{winterwerp1999dynamics} introduced additional flocs-related processes such as flocculation, gelling, and consolidation, which improved agreement with experimentally measured settling velocity of mud flocs. The volume fraction $\phi$ in the context of flocs or porous particles is defined using the volume enclosed by the outer boundary of the flocs (envelope volume), neglecting their internal porosity. Thus, for porous particles $\phi$ is defined as the ratio of the total envelope volume of all porous particles to the total volume of the system.

Although there are several models that predict the settling behavior of sediments, determining accurate sediment properties remains challenging. \citet{metelkin2025parameters} demonstrated that measurements based only on floc size and settling velocity may not be sufficient to uniquely determine the physical properties of aggregates. In their work, the authors explored the parameter space of flocculated aggregates using experimental datasets of \citet{gillard2019}, \citet{ye2020oil}, \citet{gibbs1985estuarine}, and \citet{manning1999laboratory}. The analysis showed that the relation between settling velocity and floc structure is not uniquely constrained and admits multiple combinations of parameters that yield results that are consistent with observations. In particular, it is possible to identify floc characteristics that correspond either to aggregates that are nearly impermeable with relatively low porosity, or to aggregates that are permeable with relatively high porosity. In the latter case, the effect of increased permeability is balanced by a smaller floc mass. As demonstrated in \citet{metelkin2025parameters}, these two classes of aggregates exhibit the same settling velocity when considered as isolated porous particles in an unbounded flow, yet a preliminary test case of settling within a suspension of similar aggregates showed distinct differences in the collective settling velocity of flocs. For $\phi = 0.155$, more permeable and lighter flocs exhibit a higher collective settling velocity than their less permeable, but denser, counterparts. 

The findings of \cite{metelkin2025parameters} motivate the present study, where we investigate the combined influence of floc permeability and volume fraction on the settling behavior of suspensions with varying $\phi$ by means of particle-resolved direct numerical simulations.
%To our knowledge, this is the first study that models the flocs as permeable structures, 
%\bv{BV: Are you sure that this is true? Also, is it really necessary to mention this?}
%which makes it possible to reveal how variations in floc permeability modify the hydrodynamic interactions between the flow and the particles, as well as the interactions among the particles.
The present paper is structured as follows. In §\ref{sec:methodology} we introduce the relevant equations used throughout the paper, explain the parameter range chosen for the simulation campaign, and define the simulation setup in terms of physical parameters of the system, boundary, and initial conditions. In §\ref{sec:results}, we present the resulting data of settling velocity, velocity fluctuations, particle dispersion quantities, and the resulting self-diffusion coefficients of porous particles 
and explain these effects by providing insights into
%. We also provide insights into 
particle distributions within the domain 
%by performing and analyzing Voronoï tesselation based on the particle positions and constructing pairwise particle distribution maps. 
Additionally, §\ref{sec:results} includes the analysis of counter flows that appear in the fluid region due to settling. In §\ref{sec:conclusion}, we state the main results of the present simulation campaign and highlight possible directions for future work.

\section{Computational method}
\label{sec:methodology}
\subsection{Governing equations}
\subsubsection{Fully coupled simulation framework}\label{sec:simulation_framework}
We perform particle-resolved direct numerical simulations based on a coupled Euler–Lagrange framework, in which the motion of each individual porous particle is captured. Each porous particle is modeled as a spherical, non-Brownian Lagrangian entity representing an individual floc in the context of sediment transport. The Eulerian part of the system, describing the flow, is computed by numerically solving the incompressible Navier–Stokes equations:
\begin{eqnarray}
\label{eq:NS}
\frac{\partial {\bm{u}}}{\partial t} + \nabla \cdot (\bm{u} \bm{u}) &=& - \frac{1}{\rho_f}\nabla p + \nu \nabla^2 \bm{u} - \bm{f}_b + \frac{\nu}{\kappa}\epsilon (\bm{u}-\bm{u}_p), \nonumber\\
\nabla \cdot {\bm{u}} &=& 0.
\end{eqnarray}
Here, $t$ corresponds to time, $\rho_f$ and $\nu$ correspond to the density of the fluid and its kinematic viscosity, respectively. The variable $\bm{u} = (u_x,u_y,u_z)^T$ is the fluid velocity that is continuously defined, and inside porous particles, it corresponds to the intrinsic (pore-space averaged) fluid velocity. The variable $p$ represents the pressure, and $\bm{f}_b$ corresponds to the pressure gradient that is applied to compensate for the flow that is induced by the total mass of the suspended porous particles. Such a correction is needed due to the use of periodic boundary conditions \citep{shajahan2020influence}. The last term in the right-hand side of \eqref{eq:NS} is an additional Darcy-type resistance term that is responsible for coupling between fluid and porous particle media. This porous medium is modeled as homogeneous and isotropic, with constant porosity $\epsilon$ and permeability $\kappa$ prescribed identically for all particles. For sediment flocs, porosity and permeability can be estimated from settling velocity models, with permeability expressed as a function of porosity. We evaluate permeability based on the Carman–Kozeny relation \citep{carman1939permeability}:
\begin{equation}  
    \label{eq:permeability}
    \kappa = \frac{D_p^2 \epsilon^3}{180(1-\epsilon)^2}  \qquad ,
\end{equation}
where $D_p$ denotes the diameter of the primary particles, which are assumed to constitute the porous medium and are not resolved explicitly, but instead enter the model through the prescribed permeability.
The variable $\bm{u}_p = (u_{p,x}, u_{p,y},u_{p,z})^T$ denotes the local velocity of the porous particle at the corresponding point in the Eulerian field and is calculated as
\begin{equation}
    \bm{u}_p(\bm{x}_e) = \bm{U}_i + \bm{\omega}_i \times \bigl(\bm{x}_e - \bm{X}_i\bigr) \quad .
\end{equation}
\definecolor{darkgray}{RGB}{64,64,64}
\definecolor{darkred}{RGB}{135,10,1}
\begin{figure}
    \centering
    \includegraphics[width=7.9cm]{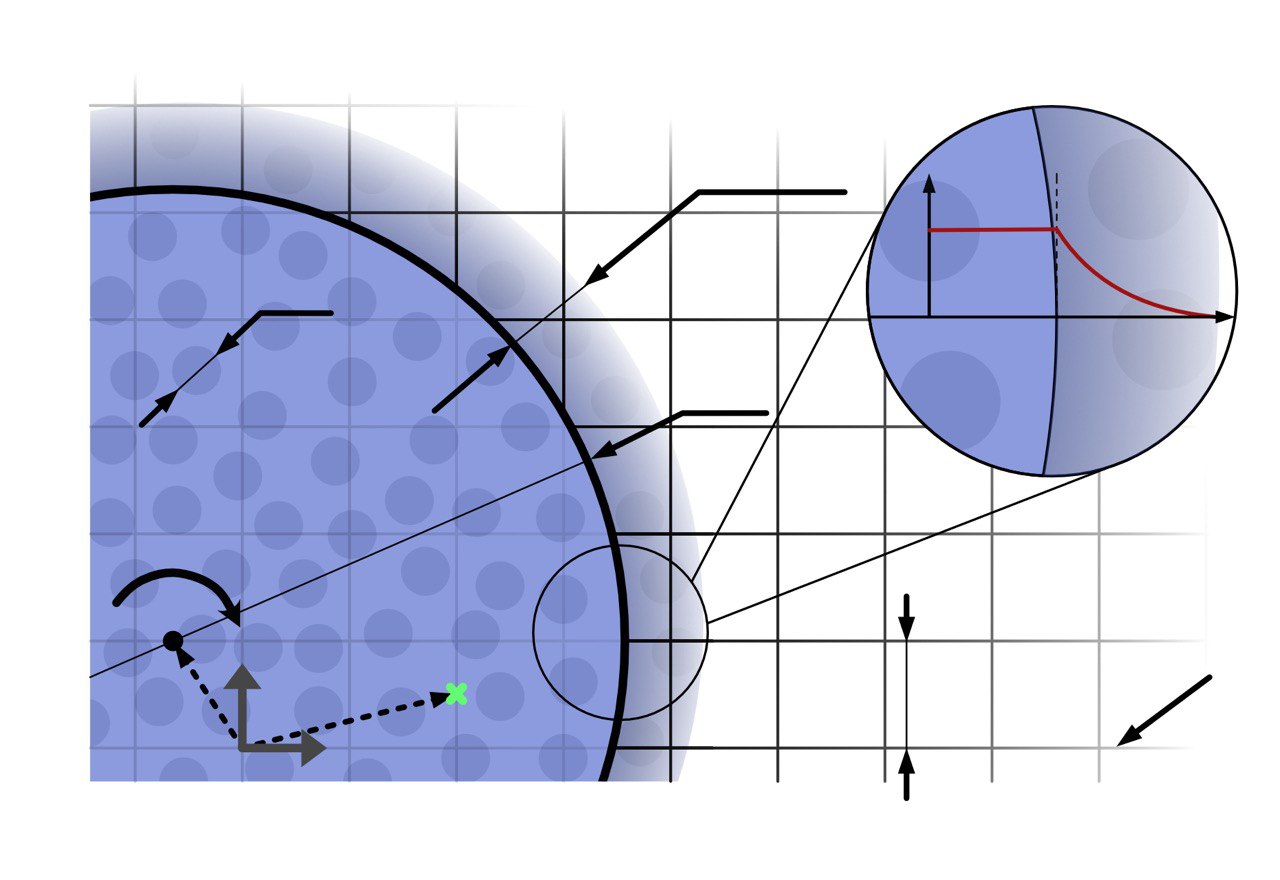}
    \put(-197,60){$\bm{\omega}$}
    \put(-101,84.5){$D$}
    \put(-100,123){$0.75h$}
    \put(-63,30){$\Delta h$}
    \put(-8,38){Eulerian grid}
    \put(-167,18){\textcolor{darkgray}{$\bm{x}$}}
    \put(-179,38){\textcolor{darkgray}{$\bm{y}$}}
    \put(-180,104){$D_p$}
    \put(-200,30){$\bm{X}$}
    \put(-160,35){$\bm{x}_e$}
    \put(-20,90){$\bm{x}$}
    \put(-60,122){\textcolor{darkred}{$\epsilon \kappa ^{-1}$}}
    \caption{Two-dimensional illustration showing the submerged porous particle within the ambient }
    \label{fig:floc_sketch}
\end{figure}
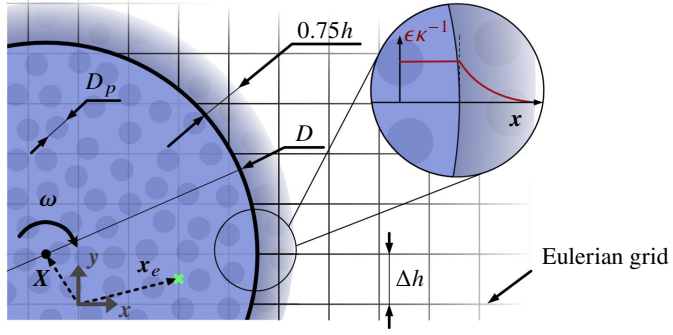
Here, $\bm{U}_i = (U_{i,x},U_{i,y}, U_{i,z})^T$ denotes the translational velocity of the particle center of mass and $\bm{\omega}$ its angular velocity. The position vector $\bm{x}_e$ refers to an arbitrary Eulerian point, while $\bm{X}_i = (X_i,Y_i,Z_i)^T$ denotes the position of the particle center. \am{The subscript $i$ denotes quantities associated with the $i$'th particle in the domain, where $i \in \{1,\dots,n_p\}$ and $n_p$ is the total number of particles.} A two-dimensional sketch of a porous particle within the computational domain is shown in figure \ref{fig:floc_sketch}, where the Eulerian point at which $\bm{u}_p$ is indicated by a green cross.
%\bv{Another word on nomenclature: all quantities relating to particles need a superscript or subscript $p$, because they relate to particle with index $p$ only. This includes velocity, position, mass, moment of inertia, forces and torques etc.}
%\am{I addressed this using subscript i - I think this is slighlty better because it makes it consistent with summation formulas, e.g. \ref{eq:pairwise_distribution} and avoids similar notation with $D_p$ where $p$ is refered to primary particles}

The coupling between the porous particles and the fluid phase implies the particle diameter to exceed the Eulerian grid spacing $\Delta h$, such that the flow within and around each particle is directly resolved. Each porous particle is assumed to be spherical with diameter $D$ and to enclose a porous medium. At the interface between a porous particle and the surrounding clear fluid, the porosity is smoothly varied, transitioning linearly from the particle porosity at the particle boundary to unity over a distance of $0.75 \Delta h$ measured normal to the particle surface and extending into the surrounding fluid (see inset in figure \ref{fig:floc_sketch}). This smoothing ensures a continuous transition between porous-medium and free-fluid regions and guarantees that, outside this interfacial layer, the Darcy term vanishes as $\kappa \to \infty$, in accordance with \eqref{eq:permeability}. The spatial variation of the porosity–permeability contribution to the Darcy term with distance from the particle center is illustrated in the zoomed-in view of figure \ref{fig:floc_sketch}.

To compute the motion of porous particles, we use the Newton–Euler equations:
\begin{eqnarray}
    \label{eq:part_dynamics}
    m_i \frac{d\bm{U}_i}{dt} &=& \bm{F}_{h,i} + \bm{F}_{g,i} + \bm{F}_{c,i} \; , \nonumber\\ 
    I_i \frac{d\bm{\omega}_i}{dt} &=& \bm{T}_{h,i} + \bm{T}_{c,i} \quad .
\end{eqnarray}
Here, the effective mass of the porous particle is expressed as $m_i = [(\rho_s(1-\epsilon) + \rho_f \epsilon] V_{agg,i}$, where $V_{agg,i}= \pi D^3/6$ corresponds to the volume occupied by a porous particle, $\rho_s$ is the density of the unresolved primary particles, and  $I_i= m_i D^2/10$ denotes its moment of inertia. The terms $\bm{F}_{h,i}$, $\bm{F}_{g,i}$, and $\bm{F}_{c,i}$ represent the respective hydrodynamic, gravitational, and contact forces, and $\bm{T}_{h,i}$ and $\bm{T}_{c,i}$ are the corresponding torques arising from fluid–particle interactions as well as collisions and contacts \am{for particle $i$}. 
%\bv{BV: give mathematical definition of $\bm{F}_{g}$.}
The hydrodynamic force is obtained from a volume integral of the Darcy term in equation \eqref{eq:NS} over the volume of the porous particle $V_{agg,i}$, namely $\bm{F}_{h,i} = \int_{V_{agg,i}} \nu \kappa^{-1} \epsilon \rho_f (\bm{u}-\bm{u}_p) dV$. The gravitational force is defined as $\bm{F}_{g,i} = \rho_s(1-\epsilon)V_{agg,i}\bm{g}$, where $\bm{g}$ is the gravitational acceleration vector. The contact force can be further decomposed into forces due to collision and contact as well as short-range hydrodynamic lubrication forces $\bm{F}_{c,i} = \bm{F}_{coll,i} + \bm{F}_{l,i}$. The contact forces comprise components in the normal and tangential directions \citep{biegert2017collision,vowinckel2021rheology}. The normal component of collision force \am{between particles $i$ and $j$} is calculated via a non-linear spring-dashpot model:
\begin{equation}
    \label{eq:Fc_norm}
    \bm{F}_{coll,ij}^n = -k_n |\zeta_n|^{3/2}\bm{n} - d_n\bm{g}_{n,cp}
\end{equation}
where $k_n$ and $d_n$ are the normal stiffness and damping coefficients that are calibrated in a way to set the restitution coefficient $e_{dry} = 0.97$. The variable $\zeta_n$ is the distance between the two contacting surfaces, the vector $\bm{n}$ is the outward-pointing normal on the contact interface, and $\bm{g}_{n,cp}$ is the normal component of the relative velocity of the particles at the contact point. The tangential component of the collision force represents a linear spring-dashpot model capped by the Coulomb friction law:
\begin{equation}
    \label{eq:Fc_tang}
    \bm{F}_{coll,ij}^t = min(-k_t \bm{\zeta}_t - d_t \bm{g}_{t,cp}, ||\mu_f \bm{F}^n_{coll,ij}||\bm{t} ) \quad .
\end{equation}
Here, $k_t$ and $d_t$ denote the tangential stiffness and damping coefficients, respectively, calculated according to \citet{thornton2011investigation,thornton2013investigation}. The variable $\bm{g}_{t,cp}$ is the tangential component of the relative velocity of the particles at the contact point. The parameter $\mu_f$ is the friction coefficient and is set to $\mu_f = 0.15$. The vector $\bm{\zeta}_t$ represents the tangential displacement accumulated over the time interval during which two particles remain in contact \citep{thornton2013investigation}, and $\bm{t}$ is the unit vector in the tangential direction. While this type of contact modeling has been verified in numerous studies for solid, rigid particles in various flow configurations \citep{biegert2017collision,vowinckel2019settling,vowinckel2021rheology,zhu2022grain}, the short-range hydrodynamic lubrication forces between particles may differ for porous particles. As demonstrated by \citet{Reboucas2021b,Reboucas2021a}, particle permeability alters the lubrication interaction, and this effect is therefore discussed separately in § \ref{sec:lub_force}.

\subsubsection{Lubrication force}
\label{sec:lub_force}
The lubrication model adopted here was originally derived analytically by \citet{cox1967slow} for a rigid spherical particle approaching a wall in Stokes flow. It was later extended by \citet{biegert2017collision} to describe interactions between spherical particles of different sizes and implemented in particle-resolved simulations of impermeable spheres as:
\begin{equation}
\label{eq:lubrication_force}
     \bm{F}_{l,ij}^s = - \frac{6\pi \mu R_{eff}^2}{max(\zeta_n,\zeta_{n,min})}\bm{g}_{n,cp} \qquad .
\end{equation}
Here, the superscript $s$ refers to solid (or impermeable) particles, $R_{eff} = R_iR_j/(R_i+R_j)$ denotes the effective radius of a particle pair $i$ and $j$, which depends on the radii of the two particles and reduces to $R_{\mathrm{eff}} = R/2$ for monodisperse spheres, and  $\mu$ is the dynamic viscosity of the ambient fluid. Full details on the definitions of these quantities can be found in \cite{biegert2017collision}. The lubrication force model is turned on for inter-particle distances $\zeta_n \leq 2 \Delta h$. The lubrication force increases as two spheres approach each other, due to the increase in pressure between particle surfaces, which arises because the fluid in the immediate proximity of the gap must be squeezed out, generating a resistance that becomes singular as the gap width tends to zero. This physical mechanism leads to the scaling of the lubrication force given by $\bm{F}_{l,ij}^s \sim 1 / \zeta_n$. While this formulation leads to $\bm{F}_{l,ij}^s \to \infty$ with $\zeta_n \to 0$,  \cite{costa2015collision} and \cite{biegert2017collision} have proposed to use $\zeta_{n,min}$ and change the denominator to $max(\zeta_n,\zeta_{n,min})$ to avoid the singularity in the lubrication force. 
Therefore, in the denominator of \eqref{eq:lubrication_force}, the parameter $\zeta_{n,min}$ is related to the length of the microtexture on the surface of the particle. 

For permeable particles, the pressure field that forms between particles is reduced. This is due to the fact that some of the fluid escapes through the particles themselves, instead of being entirely forced sideways as in the case of solid particles.
\cite{Reboucas2021a} derived a lubrication force model that depends on the permeability of porous particles. The authors took into account the assumption of small permeabilities $\mathcal{K}^{1/5} \ll 1$, where $\mathcal{K}$ is the permeability factor defined as $\mathcal{K} = \kappa /R_{eff}^2$. Using this assumption, the authors showed that for $\zeta_n > R_{eff} \mathcal{K}^{2/5}$, the lubrication force for permeable particles $\bm{F}_{l,ij}^p$ reduces to the scaling $\bm{F}_{l,ij}^p \sim \mu R_{eff}^2 \zeta_n^{-1} \bm{g}_{n,cp}$, similar to those of solid particles, i.e. \eqref{eq:lubrication_force}.
However, for $\zeta_n \leq R_{eff} \mathcal{K}^{2/5}$, the lubrication force becomes independent of $\zeta_n$ and follows the scaling
\begin{equation}
\label{eq:porous_lub_scaling}
    \bm{F}_{l,ij}^p \sim \mu R_{eff} \mathcal{K}^{-2/5} \bm{g}_{n,cp} \quad .
\end{equation}

Using numerical simulations, \cite{Reboucas2021b} provided evidence that the effect of permeability can be expressed through an equivalent roughness parameter, which is estimated as 
\begin{equation}
\label{eq:zeta_porous}
    \zeta_{n}^{p} = 0.72 \mathcal{K}^{2/5}R_{eff} \quad .
\end{equation}

Taking this into account, we use the formulation of the lubrication force as expressed in equation \eqref{eq:lubrication_force}, modified according to \eqref{eq:zeta_porous} :

\begin{equation}
    \label{eq:Lubrication_force_permeable}
    \bm{F}_{l,ij}^p = - \frac{6\pi \mu R_{eff}^2}{max(\zeta_n,\zeta_{n}^p)}\bm{g}_{n,cp}
\end{equation}

Hence, to model the lubrication force for porous particles, we use  \eqref{eq:Lubrication_force_permeable}. The corresponding values of $\zeta_{n}^p$ for different permeability values are stated later in table \ref{tab:simulation_parameters}. We keep the convention of the original lubrication model \eqref{eq:lubrication_force}, to activate the \eqref{eq:Lubrication_force_permeable} only for $\zeta_n \leq 2 \Delta h$.

%\sout{\am{For sufficiently high particle permeabilities, the minimal gap parameter $\zeta_{n,\min}$ may remain above the threshold for which the lubrication model becomes active (i.e. the regime $\zeta \leq 2h$). In this situation, the lubrication force is treated as gap-independent and is prescribed solely in terms of the fluid properties and the relative velocity between the particles.}}
%\bv{BV: Do you have any validation that this is a good model technique? Discuss the permeability range you want to investigate and point to table 1 for reference. Compare the values to those by Biegert et al. and discuss how this might affect the rule that lubrication becomes active for $\zeta_n<2h$}

\subsubsection{Non-dimensionalization of the simulation framework}\label{sec:nondimensionalisation}
We introduce non-dimensional variables denoted by the tilde operator, which modifies variables as:

\begin{eqnarray}
    \bm{x} = D \tilde{\bm{x}}, \quad
    \bm{u} = u_{st} \tilde{\bm{u}}, \quad
    \bm{u}_p = u_{st} \tilde{\bm{u}}_p ,  \nonumber\\
    t = t_{ref} \tilde{t} \quad
    p = \rho_f u_{st}^2 \tilde{p}, \quad
    \bm{f}_b = \rho_f u_{st}^2D^3 \tilde{\bm{f}_b} \quad,
    \label{eq:Non_Dim_ref}
\end{eqnarray}
where $\bm{x}$ is the location vector, $t_{ref} = D/u_{st}$ is the reference time scale and $u_{st}$ is the reference velocity scale based on the Stokes settling velocity of a single particle in creeping flow \citep{clift2005bubbles} augmented with the drag reduction factor $\Omega$ in the denominator:
\begin{equation}
    \label{eq:us_stokes}
    u_{st} = \frac{R_{agg} D^2}{18\nu \, \Omega} \quad,
\end{equation}
Here, $R_{agg}= g(\rho_{agg}-\rho_f)/\rho_f$ is the submerged specific gravity of a porous particle 
%
%\bv{BV: when you call it specific gravity,the definition should be $R_p= (\rho_{agg}/\rho_f-1)g$}
%\am{AM: I adopted definition from strom2011. Authors there define $R_p$ the same way as I do - without $g$ constant}
%
and $\rho_{agg}= \epsilon\rho_f+(1-\epsilon)\rho_s$ is the effective density of the porous particle, where $g$ is the magnitude of the gravity vector. For impermeable particles ($\Omega = 1$), equation \eqref{eq:us_stokes} reduces to the classical Stokes settling velocity.
The drag reduction factor $\Omega$ quantifies the reduction in hydrodynamic force experienced by a porous particle relative to a corresponding impermeable reference particle in Stokes flow \citep{brinkman1947calculation,ooms1970frictional,neale1973creeping} and it is evaluated as:
\begin{equation}
    \label{eq:Omega}
    \Omega = \frac{\mathrm{Drag\:of\:permeable\:sphere}}{\mathrm{Drag\:of\:solid\:sphere}} =\frac{2\beta^2(1-\beta^{-1})}{2\beta^2+3(1-\beta^{-1})} \qquad ,
\end{equation}
where $\beta$ is the non-dimensional permeability that is calculated as:
\begin{equation}
    \label{eq:beta}
    \beta = \frac{D}{2\sqrt{\kappa}} \qquad.
\end{equation}
Based on Stokes settling velocity  \eqref{eq:us_stokes}, the Stokes Reynolds number is defined as:
\begin{equation}
\label{eq:Rey_Stokes}
    \Rey_{st} = u_{st}D/\nu \quad .
\end{equation}

Using these non-dimensional variables \eqref{eq:Non_Dim_ref}, \eqref{eq:Rey_Stokes} and \eqref{eq:beta}, within equations \eqref{eq:NS}, we can rewrite the system of Navier-Stokes equations in non-dimensional form:

\begin{eqnarray}
\label{eq:NS_nondim}
\frac{\partial {\bm{\tilde{u}}}}{\partial \tilde{t}} + \nabla \bm{\tilde{u}} \bm{\tilde{u}} &=& - \nabla {\tilde{p}} + \frac{1}{\Rey_{st}} \nabla^2 {\bm{\tilde{u}}} - \bm{\tilde{f}}_b + \frac{4\epsilon \beta^2}{\Rey_{st}} (\bm{\tilde{u}}-\bm{\tilde{u}}_p), \nonumber\\
\nabla {\bm{\tilde{u}}} &=& 0 \quad .
\end{eqnarray}
%
%The Darcy number also can be expressed via the permeability factor $K$, which is used in \cite{Reboucas2021a,Reboucas2021b}, as $Da = K/4$.
%\beta is very much related to \Omega, "Da" is related to non-dimensionalization of NS and "K" was used by \cite{Reboucas2021a,Reboucas2021b} for the lubrication model. The non-dimensional permeability can also be used to estimate Darcy number which is calculated as $Da = 1/ (4 \beta^2)$.
%
%\bv{BV: Again, do you need a marker function?}
%\am{AM: I don't need it}
 
Similarly to the Navier-Stokes and continuity equations, 
%\bv{BV: you should decide whether you define Navier Stokes as the set of mass and momentum conservation or you want to treat them separately in your nomenclature. Both are possible but you should be consistent. Right now, you are mixing conventions, which is confusing.}
we apply non-dimensionalization to the particle equations of motion \eqref{eq:part_dynamics} by using the following variables:
\begin{eqnarray}
    \bm{X}_i=D \tilde{\bm{X}}_i , \quad
    m_i = \rho_f D^3 \tilde{m}_i, \quad
    \bm{U}_i &=& u_{st} \tilde{\bm{U}}_i, \quad \nonumber\\ 
    \bm{F}_i = \rho_f u_{st}^2D^3 \tilde{\bm{F}}_i, \quad 
    I_i = \rho_f D^5 \tilde{I}_i, \quad
    \bm{\omega}_i = (u_{st}/D) &\tilde{\bm{\omega}}_i&, \quad
    \bm{T}_i = \rho_f u_{st}^2 D^3 \tilde{\bm{T}}_i \quad .
    \label{eq:Non_Dim_ref_part}
\end{eqnarray}
Combining \eqref{eq:part_dynamics} with \eqref{eq:Non_Dim_ref_part} results in their non-dimensional form:
\begin{eqnarray}
    \label{eq:part_dynamics_nondim}
    \tilde{m}_i \frac{d\bm{\tilde{U}}_i}{d\tilde{t}} =& \bm{\tilde{F}}_{h,i} + \bm{\tilde{F}}_{g,i} + \bm{\tilde{F}}_{c,i} \nonumber \\
    \tilde{I}_i \frac{d\bm{\tilde{\omega}_i}}{d\tilde{t}} =& \bm{\tilde{T}}_{h,i} + \bm{\tilde{T}}_{c,i} \qquad .
\end{eqnarray}

%\bv{BV: mention that you drop the tilde in the remainder in the article.}
%\am{AM:I actually think not to drop it. Right now I defined everything in a way, that this sentence is not needed.}

The simulation approach employed in the present study is inherited from a well established code that uses the Immersed Boundary Method (IBM) to couple the dynamics of solid impermeable particles with fluid flow. The original IBM approach has been used successfully in numerous applications \citep{biegert2017collision,vowinckel2019settling,vowinckel2019consolidation,zhu2022grain,kleischmann2024pairwise,kleischmann2025long} and has previously been extended and validated for porous particle suspensions by \citet{metelkin2025parameters}. 
%\bv{BV: spell out IBM and explain what it is.}
Additional validation is provided in Appendix \ref{appA}, where the method is qualitatively assessed by inspecting the flow field around a fixed porous particle and compared with the results of \citet{yu2012numerical} and a quantitative comparison with the experimentally observed settling velocity of a particle settling towards a wall obtained by \citet{ten2002particle} yields excellent agreement.

\subsection{Considerations for hindered settling}

Starting point for our analysis of hindered settling is the semi-empirical equation of \cite{richardson1954sedfluid}:

\begin{equation}
    \label{eq:RZ_equation}
    \overline{\langle U_\parallel \rangle} = k_s \overline{U}_{\parallel,0}(1-\phi)^{n_{rz}} \quad .
\end{equation}
Here, $\overline{\langle U_\parallel \rangle} $ is the ensemble time-averaged settling velocity of the suspension. The operator $\langle \cdot \rangle$ represents the averaging over all particles in the domain, and the bar operator is related to the subsequent time averaging in a quasi-steady well-developed stage.
%\bv{BV: this definition suggests that you first perform time average for all particles before you take the spatial average for the time averaged velocity. Is this true? You should explicitly mention sequence of averaging.}
%\am{I can mathematically prove that it does not matter which averaging to do first! But I would choose to make notation $\overline{\langle U_\parallel \rangle}$, not $\overline{\langle U_\parallel \rangle}$ cause it is more pretty}
The variable $\overline{U}_{\parallel,0}$ corresponds to the converged velocity of a single particle settling in the dilute regime, and $\phi$ corresponds to particle volume fraction. The coefficient $k_s$ is unity in the original formulation of \citet{richardson1954sedimentation}. Subsequent numerical studies, however, have shown that values in the range between 0.86 and 0.92 provide a more accurate representation for suspensions with sediment volume fraction $\phi > 0.05$,
as the particle microstructure in the dilute regime becomes increasingly anisotropic and deviates towards higher values than those predicted by the RZ equation \citep{yin2007hindered,shajahan2020influence}.
%\am{which is related to the findings mentioned earlier in §\ref{sec:introduction} that in dilute regime the original RZ equation deviate from numerical data.} 
%\am{explain that area less than 5 percent is excluded because of the artifect taht most probably comes from the tripple periodic boundary condition effect. Mention it here but without emphasizing it too much}
%\bv{I am not sure I understand. First you argue that $k_s$ deviates for $\phi>0.05$ then you explain this deviation for the dilute regime}
%\am{I am not sure how to formulate this more clearly and if it is needed. I think we need to discuss it.}
%\textcolor{purple}{Yin and Koch (2007) seem to be the first ones who discussed this deviation. According to them, deviation appears due to microstructural patterns and not to the influence of periodic boundary conditions (BC). Citation from their work - "We attribute the deviation of u* from the power law to the inﬂuence of the anisotropic microstructure, which is found in dilute suspensions and is more prevalent at higher Reynolds numbers."}
%\bv{BV: is there a physical interpretation to $k_s$?}
The variable $n_{rz}$ is an empirical parameter that depends on flow properties, such as Reynolds number $Re_{\parallel,0}$ and particle density. \cite{richardson1954sedimentation} also suggested an experimentally verified relation to estimate the value of the parameter $n_{rz}$ as a function of $Re_{\parallel,0}$, based on measurements of the settling velocity of a suspension conducted in glass cylindrical tubes:
%\bv{BV: verified by what?} \am{I am not sure I addressed it clearly now.}
\begin{eqnarray}
\label{eq:n_rz_constant_rz}
    n_{rz} = 4.35 Re_{\parallel,0}^{-0.03} \quad \text{for}& \quad  0.2< Re_{\parallel,0} <1 \\ 
    n_{rz} = 4.45 Re_{\parallel,0}^{-0.1} \quad \text{for}& \quad  1< Re_{\parallel,0} <500 
\end{eqnarray}
Later, \cite{garside1977velocity} provided another empirical relation  to compute $n_{rz}$ based on comparisons with experimental data:

\begin{equation}
\label{eq:n_rz_constant_garside}
    n_{rz} = \frac{5.1 + 0.27\Rey_{\parallel,0}^{0.9}}{1+0.1\Rey_{\parallel,0}^{0.9}} \qquad .
\end{equation}
The relation \eqref{eq:n_rz_constant_garside} defines a sigmoidal dependence on the logarithm of the Reynolds number $\Rey_{\parallel,0}$, approaching a plateau at $n_{rz}=5.1$ as $\Rey_{\parallel,0}\to 0$ and converging to $n_{rz}=2.7$ as $\Rey_{\parallel,0}\to\infty$. The transition between these two asymptotic regimes occurs over the range $1 < \Rey_{\parallel,0} < 300$, outside of which $n_{rz}$ remains within $\pm \%$5 of its limiting values.
%\bv{BV: here, you could add a discussion on what happens for $Re\rightarrow 0$, because you are dealing with low $Re$ and that will be important later in the discussion.}

To capture the flow regime based on the actual settling velocity of the suspension, we use the settling Reynolds number $\Rey_\parallel$, defined from the converged ensemble-averaged settling velocity $ \overline{\langle U_\parallel \rangle}$ of the respective simulation

\begin{equation}
\label{eq:Rey_settling_real}
    \Rey_\parallel = {\overline{\langle U_\parallel \rangle}} D/\nu \quad.
\end{equation}
In simulation cases consisting of only a single particle, we apply the notation $\Rey_{\parallel,0}$.

%\bv{The following paragraph seems disconnected and it maybe worthwhile to mention it earlier, where I placed my comment.}

%\bv{A major criticism and this subsection is that it is not at all the governing equations. Those are NSE and Newton-Euler equations. This section merely provides empirical relations to describe the settling behavior of single and individual particles. It does not provide any information on flow characteristics. Hence, the caption of the subsection is misleading. We should restructure this section along the following lines.\\
%§ 2 Computational method\\
%§ 2.1 Governing equations\\
%§ 2.1.1
%Fully coupled simulations of porous particle suspensions\\
%§ 2.1.2 Extension of lubrication model to porous particles\\
%§ 2.2 Considerations for hindered settling\\
%§ 2.3 Simulation setup
%}

\subsection{Simulation setup}

\begin{table}
    \centering
    \begin{tabular}{c|c|c|c|S|c|c|c|c}
        $\Rey_{\parallel,0}$ & $\Rey_{st}$ & $\phi$ & $L_x\times L_y \times L_z$ & $n_p$ & $\Omega$ & $\epsilon$ & $D_p/D$ & $\zeta_n^p$ \\
        \hline
        \hline
         \multirow{15}{*}{0.85} & \multirow{5}{*}{1.0}  & 0 & $(20D)^3$ & 1 & \multirow{5}{*}{0.95} & \multirow{5}{*}{0.9525} & \multirow{5}{*}{$1.61\times10^{-2}$} & \multirow{5}{*}{$2.26\times10^{-1}R$}  \\
          & &0.05& $(65D)^3$&26,224& & & & \\
          & &0.1& $(65D)^3$&52,450& & & & \\
          & &0.2& $(65D)^3$&104,899& & & & \\
          & &0.3& $(65D)^3$&157,348& & & & \\
          \cline{2-9}
          &\multirow{5}{*}{0.98} & 0 & $(20D)^3$ & 1 &\multirow{5}{*}{0.8} & \multirow{5}{*}{0.961} & \multirow{5}{*}{$4.88\times10^{-2}$} & \multirow{5}{*}{$6.23\times10^{-1}R$}  \\
          & &0.05&$(65D)^3$&26,224 & & & & \\
          & &0.1& $(65D)^3$&52,450 & & & & \\
          & &0.2& $(65D)^3$&104,899& & & & \\
          & &0.3& $(65D)^3$&157,348 & & & & \\
          \cline{2-9}
          &\multirow{5}{*}{0.98} & 0 & $(20D)^3$&1 & \multirow{5}{*}{0.7} & \multirow{5}{*}{0.966} & \multirow{5}{*}{$6.2\times10^{-2}$} & \multirow{5}{*}{$8.5\times10^{-1}R$}  \\
          & &0.05& $(65D)^3$&26,224& & & & \\
          & &0.1& $(65D)^3$&52,450& & & & \\
          & &0.2& $(65D)^3$&104,899& & & & \\
          & &0.3& $(65D)^3$&157,348& & & & \\

    \end{tabular}
    \caption{Parameter space of performed simulations.
    %\bv{BV: Do you also want to include the values for $\Rey_\parallel$ and number of particles?}
    }
    %\am{AM: I don't want to include $\Rey_\parallel$ as it is result, not an input data}
    \label{tab:simulation_parameters}
\end{table}

For our simulation campaign, we consider a triple periodic domain filled with a viscous fluid laden with porous particles settling under gravity.  The complete set of physical parameters used in the present simulation campaign is summarized in table \ref{tab:simulation_parameters}. We used a constant single-particle settling Reynolds number $\Rey_{\parallel,0} = 0.85$ and five particle volume fractions $\phi \in \{0,0.05,0.1,0.2,0.3 \}$, where $\phi \rightarrow 0$ corresponds to a single particle. The total number of particles $n_p$ corresponding to each value of $\phi$ is also specified in table \ref{tab:simulation_parameters}. Prior to the suspension simulations, particle properties were selected to match a prescribed value of $\Rey_{st}$ for all cases by increasing the porosity $\epsilon$, and thereby reducing the particle mass, for cases with higher permeability, following the approach of \citet{metelkin2025parameters}. In addition, following a posteriori single-particle settling tests, the particle porosity was again slightly adjusted for $\Omega \in \{0.8,0.7\}$ in order to recover the same value of $\Rey_{\parallel,0}$.
%\bv{Should we mention how you managed to recover the same $\Rey_{\parallel,0}$? I remember it was linked to changing the density of the solid, right?.} \am{I think this is what I say in two previous sentences. Should it be more explicit?}
%\am{explicitly mention that the mass is smaller for more permeable cases. Maybe even city the previous article}
%The Stokes Reynolds number $\Rey_{st}$ is slightly lower for $\Omega \in \{0.8,0.7 \}$ due to the fact that the porosity of the particles was slightly adjusted to closely match the settling velocity of a single porous particle of the simulation with $\Omega = 0.95$. 
%\bv{BV: are you talking about 1.0 as compared to 0.98? This seems negligible.}
%\am{AM: yes, should I just ignore it then?}
%\am{\sout{The adjustment to the lubrication model by means of the equivalent roughness parameter $\zeta_n^p$ is determined using equation \eqref{eq:zeta_porous}.}} 

To understand the impact of triple-periodic boundary conditions on velocity fluctuations, we performed a preliminary campaign of simulations with varying computational domains of sizes $L_x \times L_y \times L_z \in \{10D^3, 20D^3, 40D^3, 65D^3\}$ and analyzed their results by means of fluid velocity autocorrelation (see Appendix \ref{app:autocorrelation}). Based on this analysis, a cubical domain of size $L_x \times L_y \times L_z=(65D)^3$  is selected for the main simulation campaign, as it reproduces the vertical autocorrelation trends observed in larger domains and exhibits the closest approach to zero in the horizontal direction among all tested configurations. Further details of this investigation are presented in  §\ref{sec:vel_fluct} below.
%\bv{BV: provide a quick justification for the domain size choosen for your production runs.}

\begin{figure}
    \centering
    \includegraphics[]{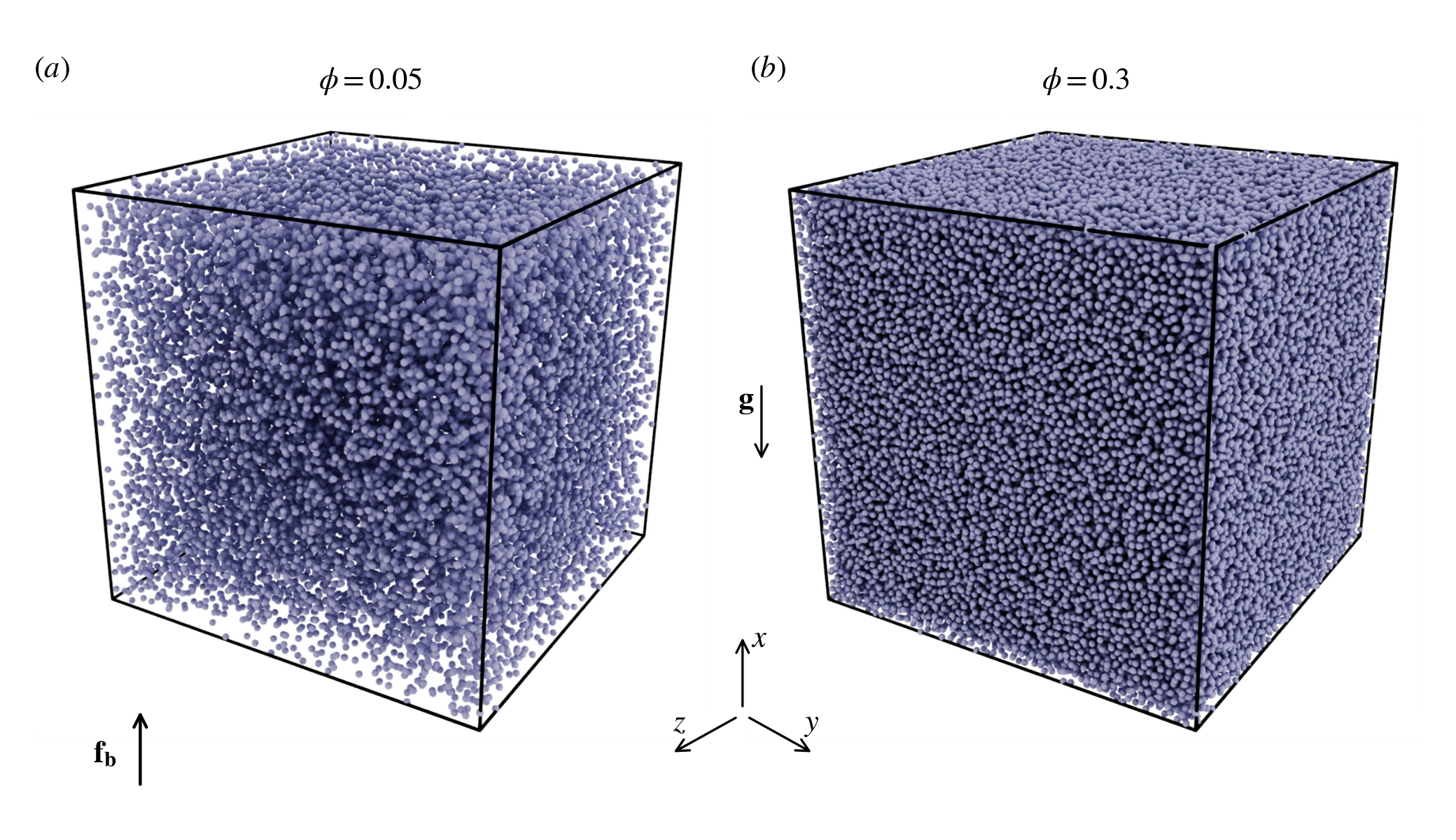}
    \caption{Illustration of simulation domain with initial distribution of porous particles for $\phi = 0.05$ $(a)$ and $\phi = 0.3$ $(b)$.
    }
    \label{fig:domain}
\end{figure}

Examples of this computational domain with the initial position of the porous particles for volume fractions of $\phi = 0.05$ and $\phi = 0.3$ are shown in figure \ref{fig:domain}. We employ triple periodic boundary conditions and apply a gravitational acceleration pointing in the negative $x$-direction, i.e. $\bm{g}=(-g,0,0)^T$. To compensate for the momentum supplied by the settling particles, we apply the negative artificial pressure gradient $\bm{f}_b = \phi(\rho_{agg}/ \rho_f-1)\bm{g}$ in \eqref{eq:NS_nondim}, which ensures zero flux for the fluid velocity in the direction of gravity. The resolution of the Eulerian grid is $N_x\times N_y\times N_z = (910)^3$, which results in a total number of grid cells of $N_{cells} = 753,571,000$.
%\bv{BV: You can be more precise}
In terms of particle discretization, this mesh resolution corresponds to 14 grid cells per particle diameter. 

Converting the system to dimensional values, the physical properties of the fluid $\rho_f = 1000\,kg/m^3$ and  $\nu = 9.2 \times 10^{-7}\,m^2/s$ correspond to water and the gravitational acceleration is set to $g=9.81\, m/s^2$. Porous particles have a fixed diameter $D = 266\,\mu\mathrm{m}$ and are composed of primary particles with density $\rho_s = 2650 \, kg/m^{-3}$, corresponding to silica. The primary-particle diameter is varied over the range $4.3\,\mu \mathrm{m} \leq D_p \leq 63\,\mu\mathrm{m}$ and, together with the porosity $\epsilon$, are adjusted to match a prescribed value of $\Omega$ and the settling velocity of an isolated particle.

%\tofocuson{For this discretization, the range of the lubrication model is equal to $2 \Delta h = 2.86 \times 10^{-1} R$, which is higher than the largest value of $\zeta_n^p = 2.38 \times 10^{-1}R$ for the most permeable particle properties. This means that even for the highest permeability values of particles, the lubrication model is gap dependent when the distance between particles $\zeta_n^p<\zeta \leq 2.86 \times 10^{-1} R$.}
%\bv{BV: comment on the spatial discretization of your porous particles and how the grid cell size does not interfere with your choices on lubrication fore modelling.}
The porous particles for all volume fractions $\phi$ were initialized by the random sequential addition algorithm that randomly positions particles without any overlap within the computational domain \citep{widom1966random}. All particles and the ambient fluid were initialized with zero velocity. We sample the output of Lagrangian particle data and Eulerian fluid data with a frequency of $0.34t_{ref}$ and  $6.8t_{ref}$, respectively. The total simulation duration, $t_{end}$, depends on the volume fraction $\phi$ and is chosen to ensure temporal convergence of all time-averaged quantities reported in §\ref{sec:results}. Simulations were performed for $t_{end} = 400 \tilde{t}$ at $\phi \in \{0.05, 0.2, 0.3\}$ and for $t_{end} = 200 \tilde{t}$ at $\phi = 0.1$ and for simulations with a single particle. 
The differences in averaging time arise from two competing effects. Simulations with fewer particles reach a quasi-steady state more rapidly, but yield larger fluctuations and fewer samples. Conversely, higher particle numbers reduce the variance but require longer to reach convergence. Consequently, simulations at $\phi = 0.05$ were run longer than those at $\phi = 0.1$, since fewer particles are available for averaging at the lower volume fraction.
%\bv{BV: give reason why you find non-monotonous dependency of $t_{end}$ on $\phi$ and}
%\bv{define how you determine $t_{end}$.}
%\sout{In the dilute limit ($\phi \rightarrow 0$), simulations were run for $t_{end} = 30 \tilde{t}$, which is sufficient to establish the well-developed state for a single particle} \am{Maybe I just remove this sentence. The meaning will not change but we won't need to exmplain those things.}
Statistical checks were performed to make sure that all quantities analyzed in this study have reached a statistically steady state within acceptable error bounds
%\bv{This seems to be a contradiction to figure 4a, where you start averaging for all simulations at $100\tilde{t}$ and run until $100\tilde{t}$. What am I missing? }

%\bv{BV: state averaging time period and point to figure 3 that clearly shows that you run simulations until you obtain a well-developed state in the bulk settling velocity.}
%\am{AM:I think it is better to state the averaging time period when discussing each plot, because they are slightly different}

\section{Results}
\label{sec:results}
\subsection{Mean settling velocity}

\begin{figure}
    \centering
    \includegraphics[]{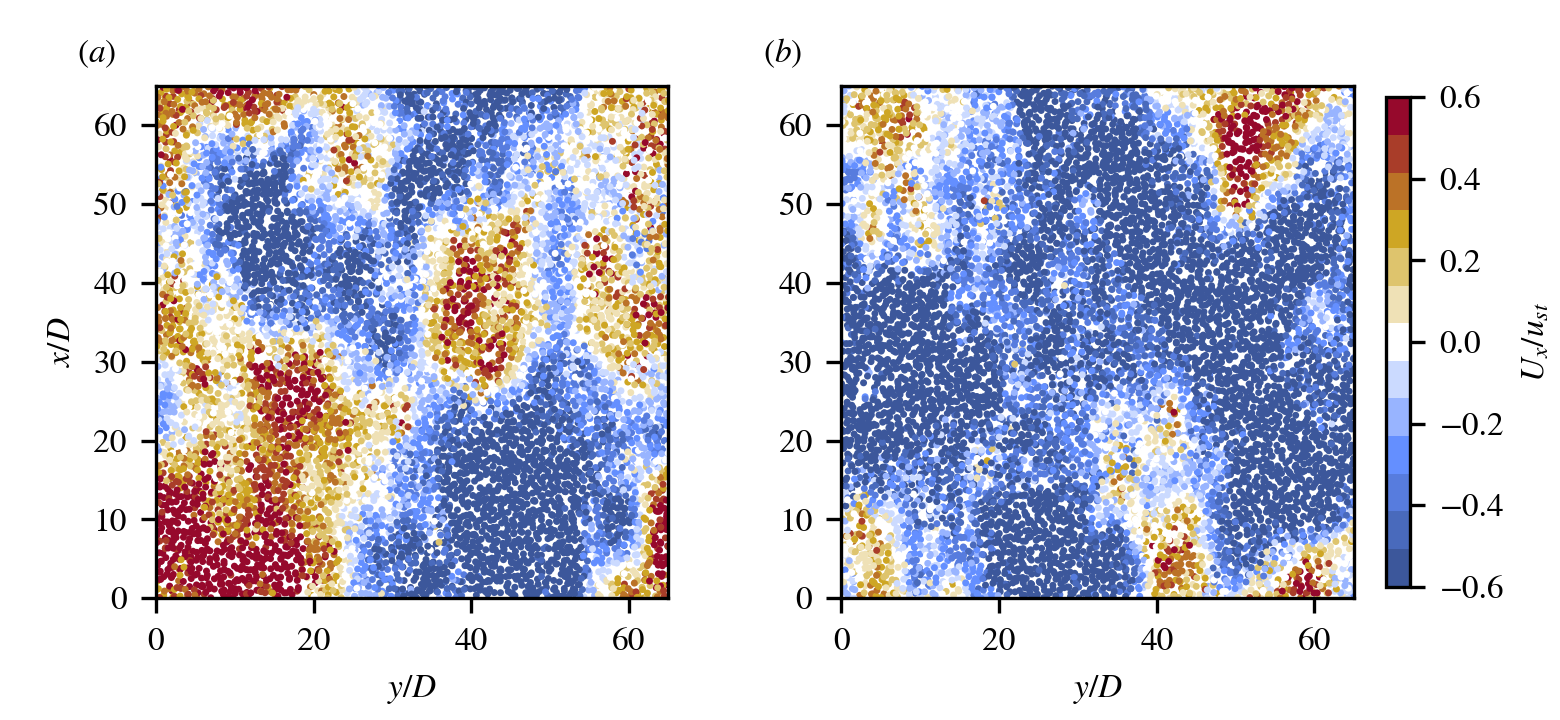}
    \caption{Side-view snapshots of the particle configurations at time $270\tilde{t}$, colored by the vertical (x-component) of the particle velocity. Panel $(a)$ shows the simulation case with $\phi = 0.3$, $\Omega = 0.95$, while panel (b) corresponds to $\phi = 0.3$, $\Omega = 0.7$.}
    \label{fig:settling_quantatively}
\end{figure}

%Description of figure 2
After initialization, the particles begin to settle under gravity because of their submerged weight. At the same time, the imposed negative artificial pressure gradient $\tilde{\bm{f}}_b$ generates an upstream flow that opposes their motion to obtain a zero net flux within the domain.
At early times, particles accelerate until the hydrodynamic drag balances the weight. Subsequently, a quasi-steady state is established, and the ensemble-averaged particle settling velocity fluctuates around its converged value. During this acceleration and its subsequent quasi-steady state, the suspension develops a heterogeneous structure: regions in which particles move downward coexist with smaller pockets where particles move upward. The emergence of the upward motion regions is governed by the definition $\tilde{\bm{f}}_b$ that requires a total mass flow rate over any horizontal plane to be equal to zero.
%These upward-moving regions arise from counterflows generated by artificial pressure gradient. 
%\bv{BV: I think you are jumping ahead with your conclusions already, because counterflows have not been evaluated yet. You can point to the fact that due to conservation of mass and momentum, you need to have regions of downward and upward moving fluids.}
A comparison between the least permeable case (figure \ref{fig:settling_quantatively}a) and the most permeable case (figure \ref{fig:settling_quantatively}b) reveals that this heterogeneity in particle motion pattern is amplified for lower particle permeability.  In agreement with observations throughout the domain and for all snapshots in the semi-steady regime, the case with $\Omega = 0.95$ exhibits a visibly larger fraction of particles advected upward compared with $\Omega = 0.7$. The reader is referred to the supplementary material for an animation comparing these two simulations over the full duration of the simulation.

%The enhanced upward transport of low-permeability particles, in turn, modifies the bulk settling characteristics of the suspension.
%\bv{BV: modifies in which way?}

\begin{figure}
    \centering
    \includegraphics[]{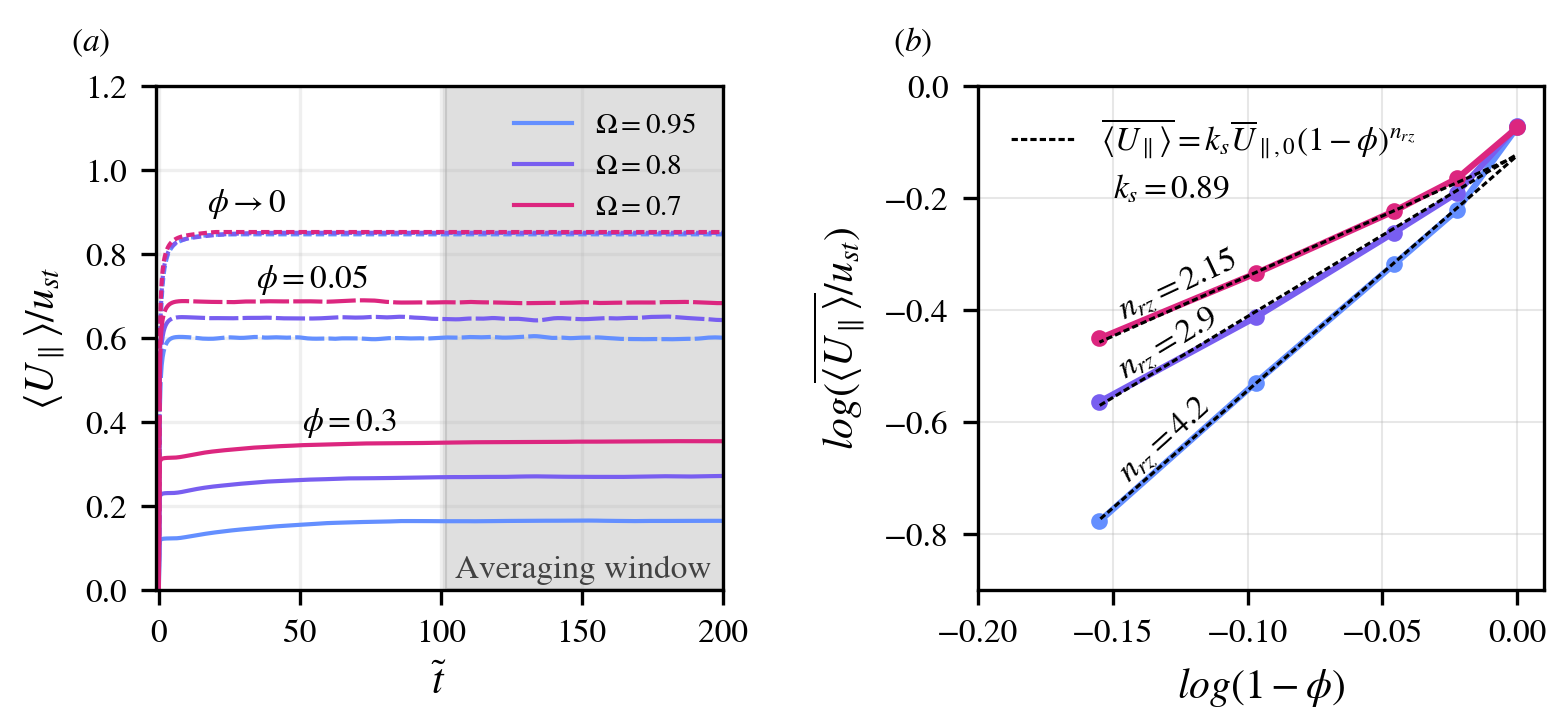}
    \caption{Ensemble-averaged settling velocity $\langle U_\parallel \rangle$ as a function of time for different $\phi$ and $\Omega$ in panel $(a)$, and time-averaged ensemble settling velocity $\overline{\langle U_\parallel \rangle}$ as a function of particle volume fraction $\phi$ in panel $(b)$.}
    \label{fig:settl_vel_over_time}
\end{figure}

%Description of Figure 3a
The primary characteristic of a suspension is the ensemble averaged settling velocity $\langle U_\parallel \rangle$. Figure \ref{fig:settl_vel_over_time}$(a)$ presents the ensemble-averaged particle velocity as a function of non-dimensional time for a selection of $\phi \in \{ 0, 0.05, 0.3 \}$ and $\Omega \in \{ 0.95, 0.8, 0.7\}$. 
%\bv{BV: Not needed - my comment was already addressed in the previous paragraph.}
%\bv{BV: that's OK, but you need to comment on the following: do they reach steady state just as fast? Do you run them for the same period of time? Is the averaging time the same?}
Figure \ref{fig:settl_vel_over_time}$(a)$ shows that the settling velocity starts from zero due to initialization and reaches a steady value over time. The duration of this transitional phase to the desired quasi-steady-state decreases with increasing particle volume fractions — ranging from around $15 \tilde{t}$ for a single particle to about $50\tilde{t}$ for $\phi = 0.3$. 
For a single particle in the domain ($\phi \rightarrow 0$), it is seen from figure \ref{fig:settl_vel_over_time}$(a)$ that the converged settling velocities are identical across all three permeability values. However, as the particle volume fraction increases, differences in the ensemble-averaged particle velocity become more pronounced. At $\phi = 0.05$, the difference between the least (dashed blue line) and most (dashed red line) permeable particles is approximately 15\%, while at $\phi = 0.3$ (blue and red solid lines) similar difference in converged settling velocity increases up to 106\%.

\begin{table}
    \centering
    \caption{Values of $\Rey_\parallel$ for the simulations of table \ref{tab:simulation_parameters} for corresponding parameters of $\phi$ and $\Omega$.}
    \label{tab:Reynolds_settling}
    \begin{tabular}{l|c c c}
         & $\Omega = 0.95$ & $\Omega = 0.8$ & $\Omega = 0.7$ \\
        \cline{1-4}
        \\[-5pt]
        $\phi \to 0$ & \multicolumn{3}{c}{$\Rey_\parallel = \Rey_{\parallel,0} = 0.85$} \\[6pt]
        $\phi = 0.05$ & 0.60 & 0.65 & 0.69 \\
        $\phi = 0.1$ & 0.48 & 0.55 & 0.60 \\
        $\phi = 0.2$ & 0.30 & 0.39 & 0.46 \\
        $\phi = 0.3$ & 0.17 & 0.27 & 0.35 \\
    \end{tabular}
\end{table}
%\bv{BV: This discussion should come after you discuss how you obtain steady-state. Hence, I recommend moving it to the bottom of this paragraph. Also, I recommend pointing to the lines you are referring to for easier assessment. I guess you mean dashed red  and dashed light blue as well as solid red and solid light blue, respectively}
%The visual comparison of simulation settling with $\phi = 0.3$ and $\Omega \in \{ 0.7,0.9 \}$ can be found in \fix{figures \ref{fig:domain}c and \ref{fig:domain}d and as animations in the supplementary material}.
%\bv{BV: give a qualitative discussion of the two new panels for figure 2 and the animations.}

%\bv{BV: Again, I recommend restructuring this paragraph. Try to be chronological. You describe that particles are released from the initial conditions with zero velocity in a quescient flow and settle under gravity, while the volume force keeps the fluid mass flux at zero. After a short initial stage that is different for different values of $\phi$ and $\Omega$ you obtain a quasi-steady state. You can also compare to Shajahan, who obtained a faster transition to steady state but he had more fluctuations in this state. Then you proceed to discuss differences in the values for the settling velocities with respect to $\phi$ as shown in figure \ref{fig:settl_vel_over_time}a.}

%Description of Figure 3b
Figure \ref{fig:settl_vel_over_time}$(b)$ shows the ensemble time-averaged settling velocity, $\overline{\langle U_\parallel \rangle}$, as a function of particle volume fraction in logarithmic coordinates. 
%\bv{BV: you need to decide whether you want to call $\overline{\langle U_\parallel \rangle}$ ensemble averaged or bulk settling velocity.}
Time averaging was performed over the interval $100 < \tilde{t} < 200$, 
%\bv{BV: is this the same for all $\phi$ and $\Omega$?}
%\am{AM: yes}
corresponding to the gray-shaded region in figure \ref{fig:settl_vel_over_time}$(a)$. Figure \ref{fig:settl_vel_over_time}$(b)$ presents $\overline{\langle U_\parallel \rangle}$ as a function of $\phi$ for different $\Omega$ together with the respective best-fit of the RZ equation as black dashed lines, along with the corresponding fitted exponents $n_{rz}$. The dependence of $\overline{\langle U_\parallel \rangle}$ on $\phi$ is well described by the RZ equation \eqref{eq:RZ_equation}, which characterizes hindered settling in suspensions. For the lowest permeability case ($\Omega = 0.95$), the fitted exponent reaches its maximum value ($n_{rz} = 4.2$) and decreases with increasing permeability, being equal to 2.9 and 2.15 for $\Omega = 0.8$ and $\Omega = 0.7$, respectively. In particular, in all cases, the RZ relation provides a good fit to the simulation results within the volume fraction range investigated $0.05 < \phi < 0.3$. The deviation between the RZ equation and the numerical results at low particle volume fractions is consistent with previous findings \citep{yin2007hindered,shajahan2020influence} and is further reflected in the fitted value $k_s = 0.89$, which lies within the range reported in the literature, $0.86 < k_s < 0.92$.
Owing to the different converged settling velocities $\overline{\langle U_\parallel \rangle}$ obtained in each simulation case, the corresponding settling Reynolds numbers $\Rey_\parallel$ differ, as reported in table \ref{tab:Reynolds_settling}. Since all other parameters defining $\Rey_\parallel$ are identical, this table provides a direct comparison of the settling velocities across the different cases.
%\bv{BV: Also comment on the value of $k_s$, which is the same for all fits and in line with previous findings.}

\begin{figure}
    \centering
    \includegraphics{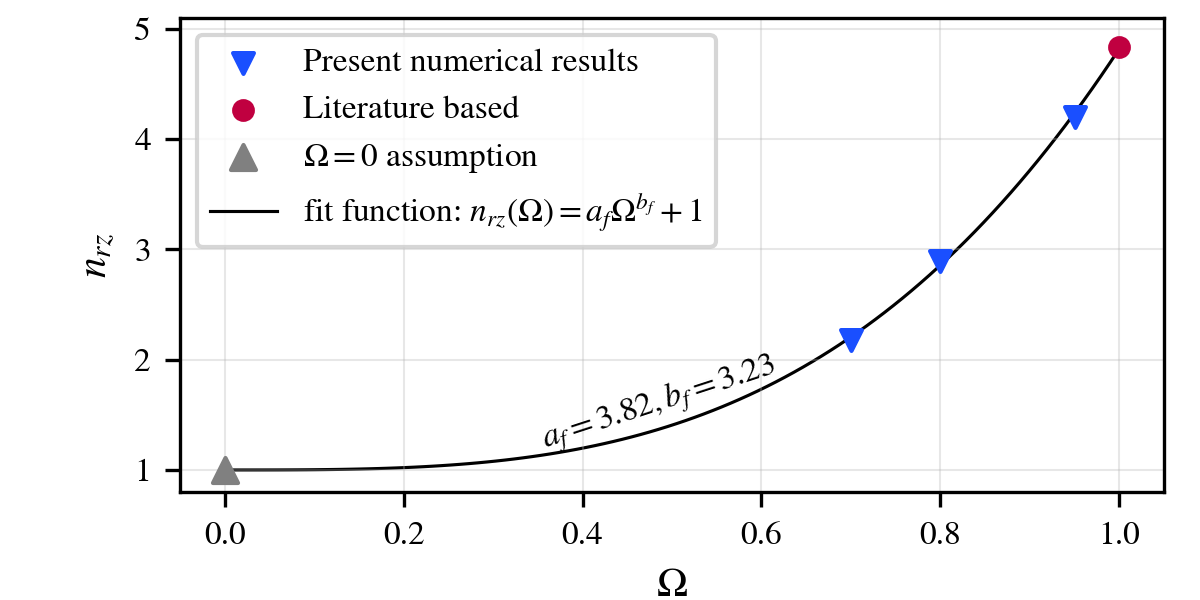}
    \caption{Exponential fit function of Richardson \& Zaki constant $n_{rz}$ as the function of drag reduction factor $\Omega$.
    }
    \label{fig:n_vs_Omega}
\end{figure}

%Description of Figure 4
%Given that decreasing $\Rey_{\parallel,0}$ causes the exponent $n_{rz}$ to approach a constant value \bv{for $\Rey_\parallel<0$}, as predicted by equation \eqref{eq:n_rz_constant_garside}, we infer that the functional dependence of $n_{rz}$ at lower $Re_{st}$ follows the same qualitative trend as observed for the present case with $\Rey_{\parallel,0} = 0.85$. 
%\bv{BV: I do not understand this sentence. Please rephrase. If I understand correctly, every simulation has a different $\Rey_\parallel$ and this would yield different $n_{rz}$. Also comment, if possible, on the critical value of $\Rey_\parallel$, where this assumption is expected to break down.}
On this basis, and in order to generalize the dependence of the exponent $n_{rz}$ on the drag reduction factor $\Omega$, we fitted a power-law. This relationship is presented in figure~ \ref{fig:n_vs_Omega}. For completeness, we supplement the data with the extreme cases of impermeable ($\Omega=1$) and fully permeable particles ($\Omega=0$). For $\Omega=1$, we find
$n_{rz} = 4.845$ 
obtained as the average 
of the predictions provided by \eqref{eq:n_rz_constant_rz} and  \eqref{eq:n_rz_constant_garside} that yield $n_{rz} = 4.37$ and $n_{rz} = 5.32$, respectively,  for $\Rey_{\parallel,0} = 0.85$.
%\bv{BV: Why do you use $\Rey_{\parallel,0} = 0.85$?}
%\am{AM: Because it corresponds to $\Rey_{st} = 1$, which I defined a priori.}
For $\Omega \to 0$ we assume $n \to 1$, since in this limit the hydrodynamic force exerted by the particle on the fluid vanishes, leading to an undisturbed flow field and thereby suppressing all mechanisms responsible for hindered settling. Consequently, the resulting fit of the form
\begin{equation}
    \label{eq:n_rz_func_omega}
    n_{rz} = a_f\Omega^{b_f} + 1
\end{equation}
closely follows the present results with fitting parameters $a_f = 3.82$ and $b_f = 3.23$. We assume that the fitting function \eqref{eq:n_rz_func_omega} remains applicable for smaller values of $\Rey_{\parallel,0}$. This assumption is motivated by relation \eqref{eq:n_rz_constant_garside}, which shows that $n_{rz}(\Rey_{\parallel,0})$ approaches a plateau for $0 < \Rey_{\parallel,0} < 1$, converging to $n_{rz}=5.1$ as $\Rey_{\parallel,0} \to 0$. On this basis, we expect \eqref{eq:n_rz_func_omega} to be applicable in the range $0 < \Rey_{\parallel,0} < 1$ using the same fitting constants, $a_f = 3.82$ and $b_f = 3.23$.

%\bv{BV: one thing that you may or may not choose to comment on: how do you expect this to change for different Reynolds numbers?}

\subsection{Counter flows}
\label{sec:counter_flows}
To better understand the reason for the faster settling of more porous particles observed in figure \ref{fig:settl_vel_over_time}b, we examine the effect of counter flows within the domain as one of the main mechanisms governing hindered settling. The permeability of the porous particles alters the return flow around them, and hence, it is of interest to evaluate the impact of particle permeability on the hindered settling behavior. To assess the effect of counterflows, we compute the average vertical velocity of the Eulerian flow field in the part of the domain that is not occupied by particles:
\begin{equation}
    \label{eq:counter_flow_velocity}
    u_{f,\parallel} = \frac{1}{V_{f}N_t} \int_{180\tilde{t}}^{200\tilde{t}}\int _{V_{f}} u_x(\bm{x},t_k) dV d\tilde{t} \quad ,
\end{equation}
where $V_f=(1-\phi)V_{dmn}$ is the volume of the simulation domain not occupied by porous particles, where $V_{dmn}$ corresponds to the total volume of the domain. 
%\am{The variable $N_t$ denotes the number of time samples taken at constant intervals of $4\tilde t$ after convergence of the settling velocity, over the range $180 \le \tilde t \le 200$, i.e. at $\tilde t \in \{180,184,188,192,196,200\}$.}
%\am{change summation operator to integral over this period of time }
%\bv{How many samples? What is the value of $N_t$?}
%\bv{BV: $N_t$ is only a good measure if samples are taken at regular time intervals.}

\begin{figure}
    \centering
    \includegraphics[]{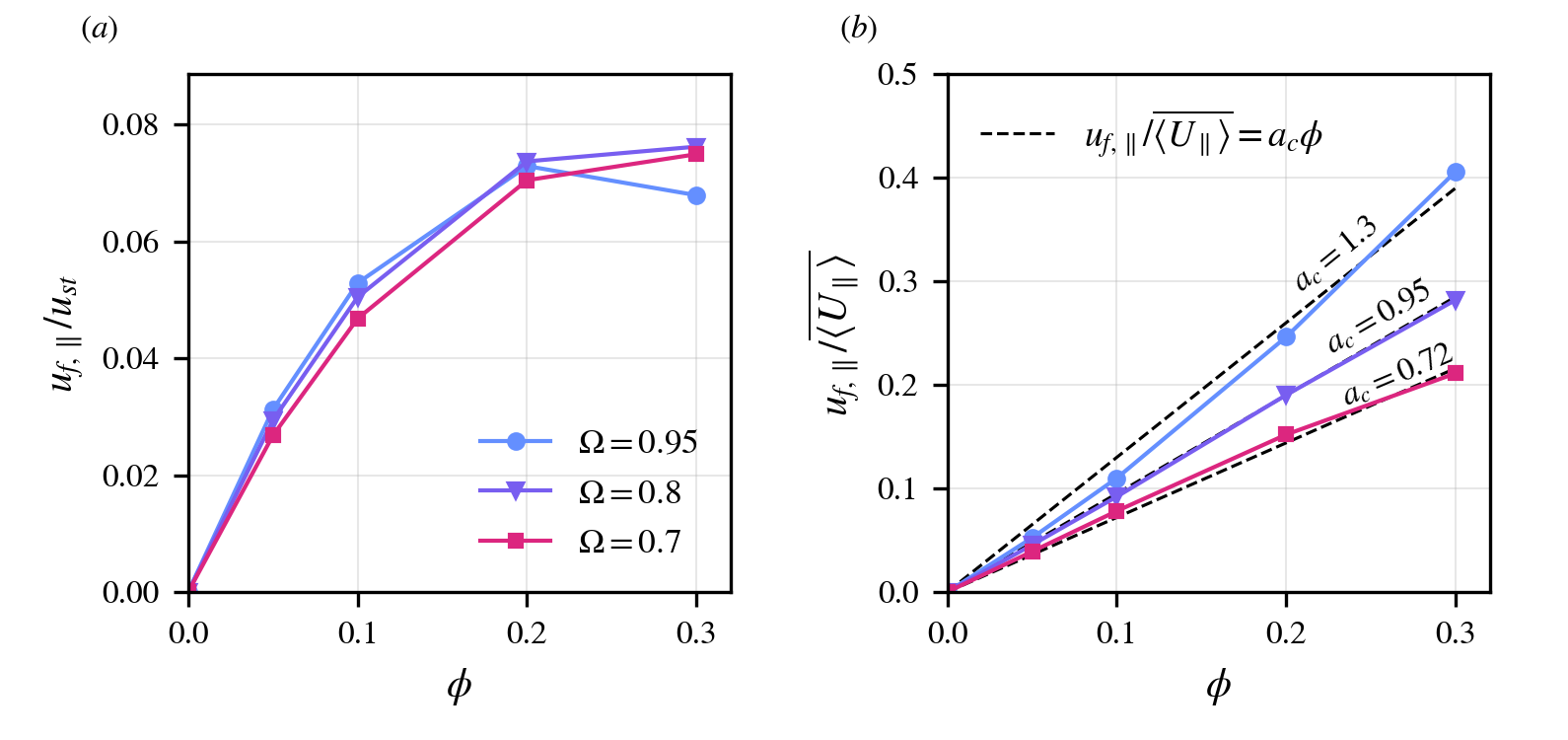}
    \caption{Counter flow velocity $u_{f,\parallel}$ as function of $\phi$ for different values of $\Omega$ normalized by the reference velocity $u_{st}$ as defined by \eqref{eq:us_stokes} in panel $(a)$ and normalized by the corresponding velocity of the suspension $\overline{\langle U_\parallel \rangle}$ in panel $(b)$.
    }
    \label{fig:counter_flows}
\end{figure}

Figure \ref{fig:counter_flows}$(a)$ shows the mean vertical fluid velocity ($u_{f,\parallel}$) normalized by the reference velocity $u_{st}$. The results show that $u_{f,\parallel}$ increases with the particle volume fraction $\phi$ to approximately $\phi = 0.2$ for all values of $\Omega$. At $\phi = 0.3$, the behavior diverges: the velocity saturates for $\Omega \in \{0.7,0.8\}$, whereas it decreases for $\Omega = 0.95$. Despite these differences, the absolute values of the counterflow velocities remain broadly comparable for the different values of $\Omega$. An initial increase of $u_{f,\parallel}$ followed by a plateau or a decline can be understood as the result of following two competing mechanisms. On the one hand, increasing $\phi$ reduces the volume of free fluid and, consequently, the effective cross-sectional area available for upward movement. For a given particle settling velocity $\overline{\langle U_\parallel \rangle}$, this geometric restriction requires an increase in fluid counterflow to satisfy mass conservation. At sufficiently high $\phi$, on the other hand, we find the averaged particle settling velocity $\overline{\langle U_\parallel \rangle}$ to decrease (see figure \ref{fig:settl_vel_over_time}), which can be attributed to an increasing number of frictional contacts between the particles. This, in turn, diminishes the momentum transferred from the particles to the fluid phase that is not occupied by porous particles, thus decreasing the counterflow velocity. The balance between these opposing effects explains the non-monotonic dependence observed in figure \ref{fig:counter_flows}$(a)$. As $\phi$ increases further, the settling velocity, and hence the velocity of counterflows, is expected to approach zero as the suspension enters a jammed regime in which particle motion is strongly restricted. Since such volume fractions are not considered here, this regime is not analysed in the present work.
%As $\phi$ increases further, the settling velocity steadily decreases and eventually approaches zero as the suspension enters a jamming regime where particle motion is strongly restricted. In this limit, the counterflows produced by particle-induced fluid displacement also disappear, reflecting the absence of any sustained particle settling.
%\bv{BV: you need to emphasize that you have not analyzed the jamming regime, but you are speculating what would happen if you increase $\phi$ further based on empirical evidence from previous studies.}

Although figure \ref{fig:counter_flows}(a) shows that the absolute values of counterflow velocity $u_{f,\parallel}/u_{st}$ remain similar for different $\Omega$ at a fixed $\phi$, the settling velocity of the suspension varies significantly with permeability. To highlight this difference, figure \ref{fig:counter_flows}$(b)$ presents the counterflow velocity normalized by the settling velocity of the suspension as a function of $\phi$. The results reveal an approximately linear dependence, with the slope increasing for higher values of $\Omega$ as indicated by the linear fits (dashed lines) and its corresponding slopes $a_c$. At $\phi=0.3$, the relative counterflow velocity differs by nearly a factor of two between the lowest and highest permeability, indicating a stronger momentum transfer to the fluid as $\Omega$ increases. Thus, the data clearly demonstrate that suspensions composed of less permeable particles (higher $\Omega$) generate counterflows with a higher relative upward velocity. This finding provides quantitative support for the hypothesis originally proposed by \citet{metelkin2025parameters}, who suggested that counterflows intensify with decreasing particle permeability. The present results not only confirm this observation but also provide quantitative evidence on how permeability modulates the efficiency of momentum transfer and the resulting fluid motion.

%\sout{However, it should be noted that the observed differences in counterflow velocity cannot be directly attributed to variations in the effective particle mass or its solid fraction. For example, the ratio of the effective particle mass $m$ between the cases with $\Omega = 0.95$ and $\Omega = 0.7$ amounts to only $2.1\%$, while the corresponding ratio of the mass of the solid fraction of particle, $m_{s} = V \rho_s (1 - \epsilon)$, is approximately $40\%$. By comparison, the maximum ratio of the absolute counterflow velocity $u_{f,\parallel}$ between $\Omega = 0.95$ and $\Omega = 0.7$ reaches $17\%$.}
%\bv{BV: This is a confusing arguement and I do not understand. We should discuss or you find a way to simplify. }
%\sout{When normalized by the suspension settling velocity, the counterflow velocity exhibits a maximal ratio of $92\%$.}
%\bv{BV: where do I see this?}
%\sout{These results highlight that permeability-driven changes in hydrodynamic interactions are the dominant mechanism underlying the observed variations in counterflow intensity.}

\subsection{Velocity fluctuations}
\label{sec:vel_fluct}
%Although the numerical simulations presented here are a promising and convenient tool to investigate the well-developed state of hindered settling, it must be noted that velocity fluctuations have been found to correlate with the size of the computational domain if triple periodic boundary conditions are applied. \am{With those boundary conditions, velocity fluctuations grow as the domain increases in size, which} may present an artefact of the boundary conditions \citep{ladd2002effects,yin2008velocity, guazzelli2011fluctuations}. 
%\bv{BV: you should point out that these fluctuations grow when you increase your domain size and that the observations in those references were for solid particles, but it is unknown how this might change for porous particles.}
%\am{AM: I would choose not to mention the point that for porous particles this correlation does not hold. Because there are no reasons why it should not. Moreover, I don't discuss this further in the text, for example, I don't have results similar to those in figure \ref{fig:u_fl_over_L} for different $\Omega$}

Velocity fluctuations are closely linked to the emergence and scale of coherent structures in settling suspensions and serve as an important indicator of collective hydrodynamic behavior. Together with particle dispersion, they form the principal statistical measures used to characterize suspension mixing. Therefore, understanding how particle permeability affects the magnitude, anisotropy, and volume fraction dependence of velocity fluctuations is essential for quantifying its impact on the collective dynamics of settling suspensions.

In the discussion of velocity fluctuations, it is important to first contextualize the results with respect to experimental observations. According to the review by \citet{guazzelli2011fluctuations}, experimentally measured velocity fluctuations depend on the container size only when the container is small relative to the particle size. Specifically, this applies for containers where the minimum dimension $L_{min}$ is smaller than $L_{min} < 10D \phi ^{-1/3}$. For $\phi = 0.05$ and $\phi = 0.3$, the value of the minimum dimension of the container is $L_{min} \approx 27D$ and $L_{min} \approx 15D$, respectively. However, for containers larger than this threshold, the velocity fluctuations become independent of the container size. 
%In experiments for the size-independent regime, \citet{guazzelli2011fluctuations} show that at $\phi = 0.3$ the vertical and horizontal velocity fluctuations peak at approximately $1.7$ and $1.1$ times the mean settling speed, respectively.
%\bv{BV: provide a linkage to our simulations. Based on this explanation, this would mean our container should be big enough if we didn't use periodic boundary conditions, right?}
%\am{AM: This comment I didn't understand. The next paragraph discusses the size of the container with walls BC}

In contrast, simulations employing triple periodic BC show that velocity fluctuations scale with the container size regardless of the domain size \citep{ladd2002effects,cunha2002modeling,yin2008velocity}. These studies showed that velocity fluctuations scale linearly with domain size in the viscous limit $\Rey_\parallel \to 0$ and logarithmically at intermediate Reynolds numbers $1 < \Rey_\parallel < 10$. Despite this limitation, triple periodic BC allow velocity fluctuations to reach convergence at a substantially lower computational cost and hence, those BC are usually chosen to study converged settling of particles \citep{uhlmann2014sedimentation,yao2021effects,shajahan2023inertial}. With horizontal walls, particles accumulate at the bottom, and the upper region becomes depleted, reducing the number of particles available for ensemble averaging at the prescribed volume fraction. Moreover, achieving statistical convergence would require a domain sufficiently extended in the settling direction to allow particles to experience fully developed flow before deposition. For these reasons, triple periodic boundary conditions are adopted here to enable long-time simulations and reliable statistics in a computationally efficient manner.
%\bv{BV: as you pointed out, we should insist on the fact that it is not only us, but many other researchers do this.}
%\bv{BV: This reads very well in principle, but I would like you to consult ChatGPT to shorten this paragraph as it comes along as very verbose.}

\begin{figure}
    \centering
    \includegraphics{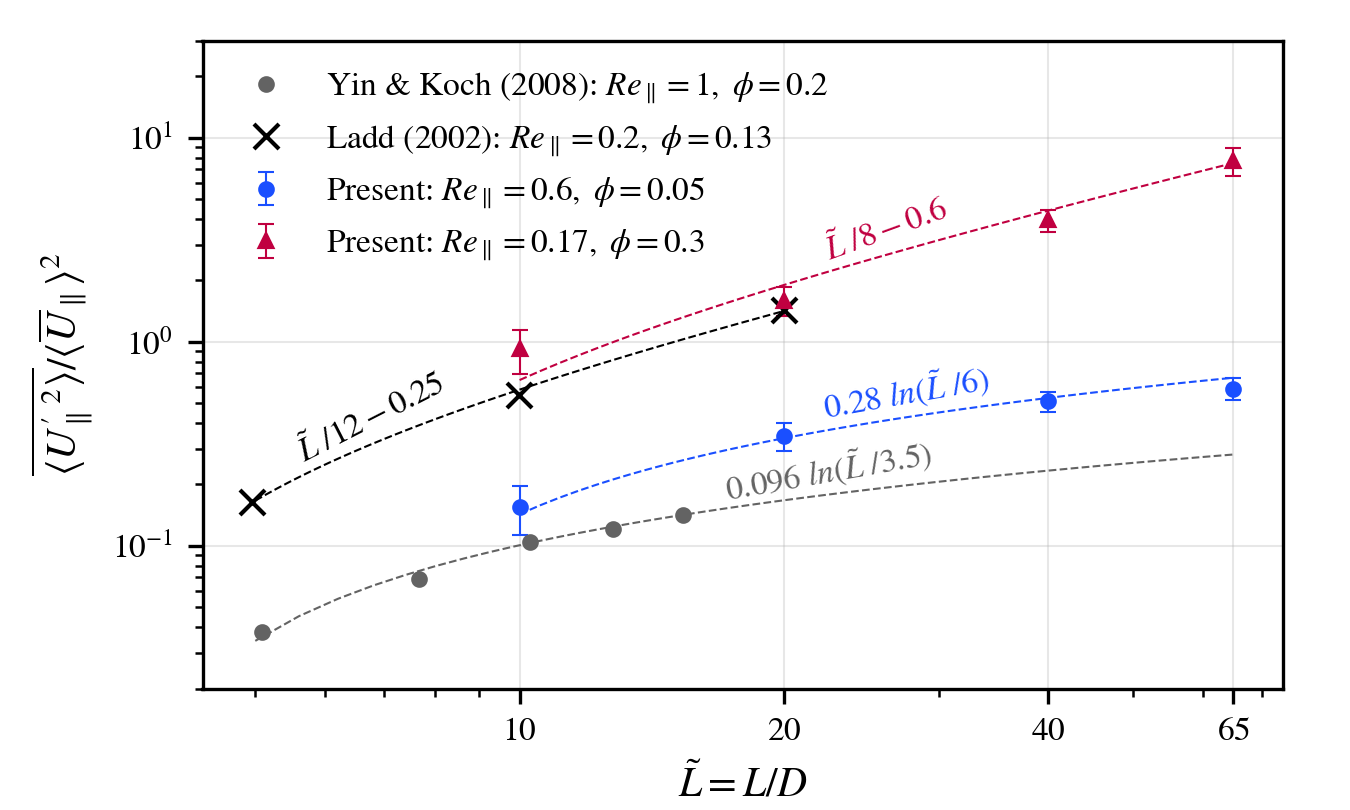}
    \caption{Velocity fluctuations as a function of domain size for $\Omega = 0.95$ from table \ref{tab:simulation_parameters}. Error bars indicate the standard deviation from time mean. Dashed lines indicate the best-fit curves for the respective linear or logarithmic relations.
    %\bv{BV: Change labeling to $\overline{U}$.}
    }
    \label{fig:u_fl_over_L}
\end{figure}

Since the computational domain used in the present simulations is larger than that employed in studies investigating similar settling regimes \citep{ladd2002effects,yin2008velocity}, higher velocity fluctuations are expected. To verify that our simulations reproduce the same dependence of velocity fluctuations on domain size, we performed additional simulations for particle  $\phi \in \{0.05, 0.3\}$ and using the almost impermeable type of particle ($\Omega=0.95$). As reported in table \ref{tab:Reynolds_settling}, the settling Reynolds numbers are $\Rey_\parallel = 0.6$ for $\phi = 0.05$ and $\Rey_\parallel = 0.17$ for $\phi = 0.3$. The settling velocity is found to be independent of the domain size, and accordingly, the settling Reynolds number does not vary with domain size.

Figure \ref{fig:u_fl_over_L} shows the time averaged mean squared fluctuations of the vertical particle velocity over the domain size defined as
\begin{equation}
    \label{eq:velocity_fluct}
    \overline{\langle {{U'}^2}_\parallel \rangle} = \overline{\langle {{U_x'}^2} \rangle} = \frac{1}{t_{end}-t_{fs}} \int_{t_{fs}}^{t_{end}} \langle(U_x(t) - \langle U_x (t) \rangle)^2\rangle dt \quad .
\end{equation}
as a function of the non-dimensional cubical domain size $\tilde{L}$. The values of $t_{fs}$ for time-averaging are equal to $100\tilde{t}$ for $\phi = 0.05$ and $200\tilde{t}$ for $\phi = 0.3$, respectively.
%\bv{BV: provide exact mathematical definition of your velocity fluctuations}
The values of $\overline{\langle {{U'}^2}_\parallel \rangle}$ are normalized by the squared mean settling velocity, $\langle U_\parallel \rangle^2$. The simulations were conducted in cubic domains of size $L_x \times L_y \times L_z \in \{(10D)^3, (20D)^3, (40D)^3, (65D)^3\}$. 
%\bv{BV: why not just call it $L$?} \am{AM:to be consistent with appendix}
The error bars in figure \ref{fig:u_fl_over_L} indicate the standard deviation of the fluctuations associated with the time averaging.
All simulations were conducted with the same fluid viscosity and particle properties \am{that correspond to a value of $\Omega = 0.95$ from table \ref{tab:simulation_parameters}}. Hence, the variations in the settling Reynolds number $\Rey_\parallel$ arise solely from differences in the ensemble-average particle settling velocity, which changes for different particle volume fractions. 
%\bv{You should briefly recap the particle properties.}

For $\Rey_\parallel = 0.6$ and $\phi = 0.05$, a logarithmic dependence of velocity fluctuations on the size of the domain is observed, consistent with the relation proposed by \cite{yin2008velocity}, $ \overline{\langle {U^\prime_\parallel}^2 \rangle} / \langle U_\parallel \rangle^2 \propto a_l \ln(\tilde{L}/b_l)$, where the fitting constants are $a_l = 0.28$ and $b_l = 6$. The magnitude of the fluctuations observed in this study is higher than those reported by \cite{yin2008velocity}. We explain this difference by the fact that in simulations of \cite{yin2008velocity}, the value of $\Rey_\parallel = 1$ is higher by comparison with our simulations where $\Rey_\parallel = 0.6$. At a higher particle volume fraction ($\phi = 0.3$) corresponding to $\Rey_\parallel = 0.17$, the velocity fluctuations are further amplified. Also, the dependence of squared velocity fluctuations on $\tilde{L}$ appears to be more linear, in agreement with the findings of \cite{ladd2002effects}, who reported a similar trend at $\Rey_\parallel = 0.2$ and $\phi = 0.13$. Therefore, as can be seen from figure \ref{fig:u_fl_over_L}, the general trend of velocity fluctuations is consistent with studies of \cite{yin2008velocity} and \cite{ladd2002effects}. These results additionally validate our numerical approach.
%\sout{and account for the substantially larger velocity fluctuations observed for the present simulations, as discussed later in this section.}} 
%\bv{"Substantially larger" does not sound good :-(  ...and it leaves the reader a bit unsatisfied. In general, you are undercutting your achievements here. You said already and prominently enough that fluctuations are a feature of the triple periodic domain. We do not need to stress it again.}
%\bv{BV: one more comment about nomenclature. Until now, it seemed you used the convention $\bm{u}=(u,v,w)^T$, but now you define your particle velocity vector with index notation $\bm{U}=(U_x,U_y,U_z)^T$. This seems inconsistent and I recommend to use $\bm{U}=(U,V,W)^T$ instead.}
%\bv{BV: Even though our fluctuations are a bit larger because of the large domain size, We can still highlight a few advantages of our domain. Those are, e.g. larger particle clusters and larger statistical samples.}

\begin{figure}
    \centering
    \includegraphics{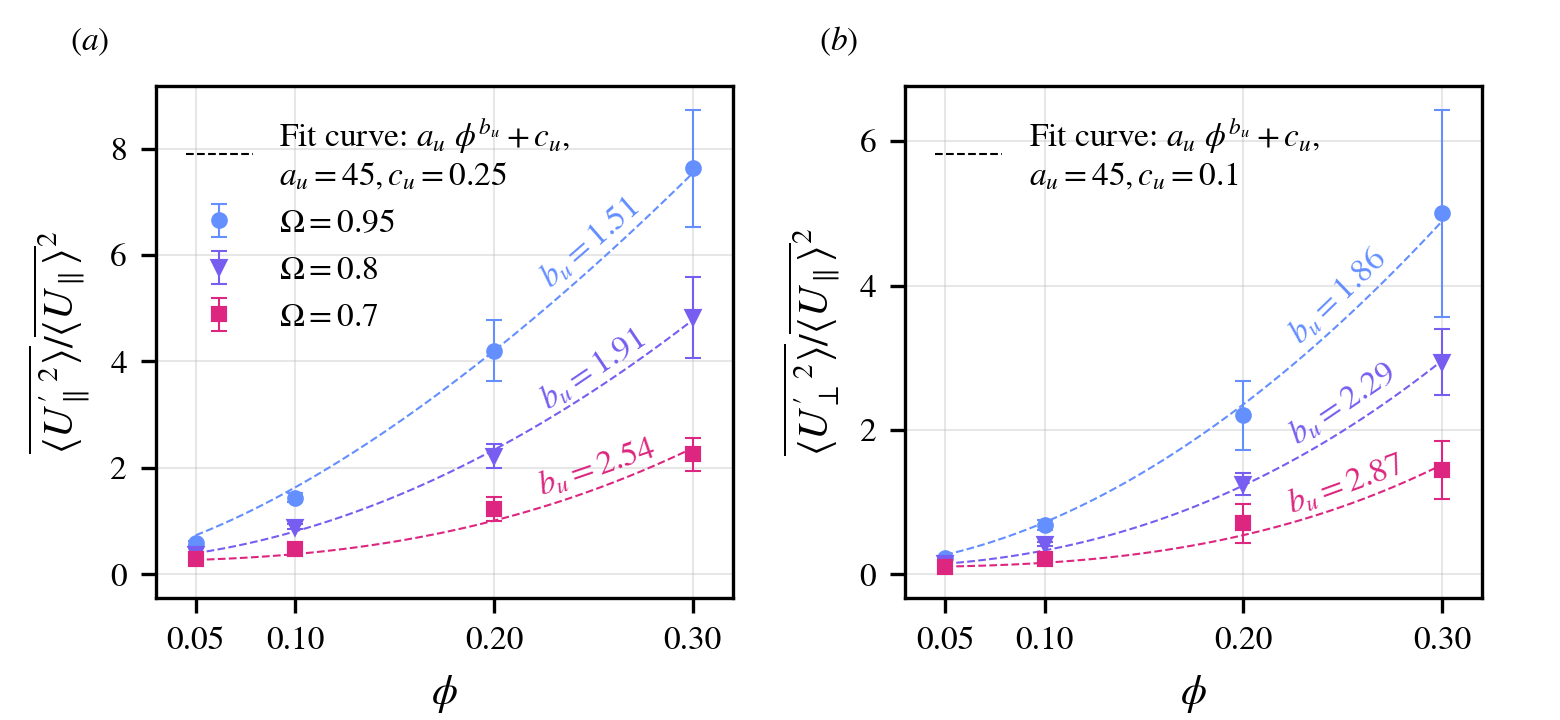}
    \caption{Vertical $(a)$ and horizontal $(b)$ velocity fluctuations over volume fraction of porous particles. Error bars indicate the standard deviation from time averaging.}
    \label{fig:u_fl_over_phi}
\end{figure}

%\bv{BV: when you start a paragraph with a sentence like that, you should provide a brief motivation why this analysis is useful.}
%\sout{Acknowledging the limitations of the numerical approach, we can use this behavior to }
We furthermore evaluate the effect of particle permeability on the velocity fluctuations. Towards this end, we show the comparison of velocity fluctuations for different values of $\Omega$ and $\phi$, with the vertical and horizontal components in figure \ref{fig:u_fl_over_phi}$(a)$ and $(b)$, respectively. 
%\bv{BV: I would not call it streamwise/spanwise (applies more to horizontal channel flow) but vertical/horizontal (better for settling).}
%\am{Done!}
The horizontal component of the averaged squared particle velocity fluctuations is defined as $\overline{\langle {U^\prime_\perp}^2 \rangle} = \overline{\langle {U^\prime_y}^2 \rangle} + \overline{\langle {U^\prime_z}^2 \rangle}$, where the components of $\overline{\langle {U^\prime_y}^2 \rangle}$ and $\overline{\langle {U^\prime_z}^2 \rangle}$ are calculated the same way as $\overline{\langle {U^\prime_x}^2 \rangle}$, (see \eqref{eq:velocity_fluct}). Error bars for each data point represent the standard deviation of the mean.
Similarly to figure \ref{fig:u_fl_over_L}, all velocity fluctuations are normalized by the squared mean settling velocity of the particles $\overline{\langle {U}_\parallel \rangle}^2$. It is observed that both fluctuation components reach their highest values for the least permeable particles ($\Omega = 0.95$) over the full range of tested volume fractions (light blue symbols and lines in figures \ref{fig:u_fl_over_phi} $(a)$ and $(b)$). It is evident that, over the full range of particle volume fractions examined, the relative velocity fluctuations systematically decrease with decreasing $\Omega$. This trend is also a consequence of the normalization by the mean settling velocity, which increases with lower $\Omega$ as seen from figure \ref{fig:settl_vel_over_time}. Moreover, velocity fluctuations systematically increase with particle volume fraction in the range $0.05 < \phi < 0.3$, regardless of permeability. Fitting a power-law curve of the form $a_{u}\phi^{b_{u}} + c_{u}$ to the data for each $\Omega$ reveals clear trends.
%\bv{BV: this is the second time you use $a$ and $b$ as fitting parameters in different context. This can lead to confusion. Try to use unique symbols for every analysis.}
The parameters $a_{u}$ and $c_{u}$, remain relatively consistent in all cases. Therefore, for the fits presented in figure \ref{fig:u_fl_over_phi}, we fixed $a_{u}=45$ for both components, with $c_{u}=0.25$ for the vertical fluctuations and $c_{u}=0.1$ for the horizontal fluctuations. The exponent $b_{u}$, however, decreases with decreasing $\Omega$, indicating stronger nonlinearity at higher permeability. 
This demonstrates once more the effect of permeability on the settling of porous particles, which allows flow through the particles, so that not only counterflows are diminished, but also fluctuations in the fluid velocity are reduced.

\subsection{Particle dispersion and self-diffusion}
\begin{figure}
    \centering
    \includegraphics[]{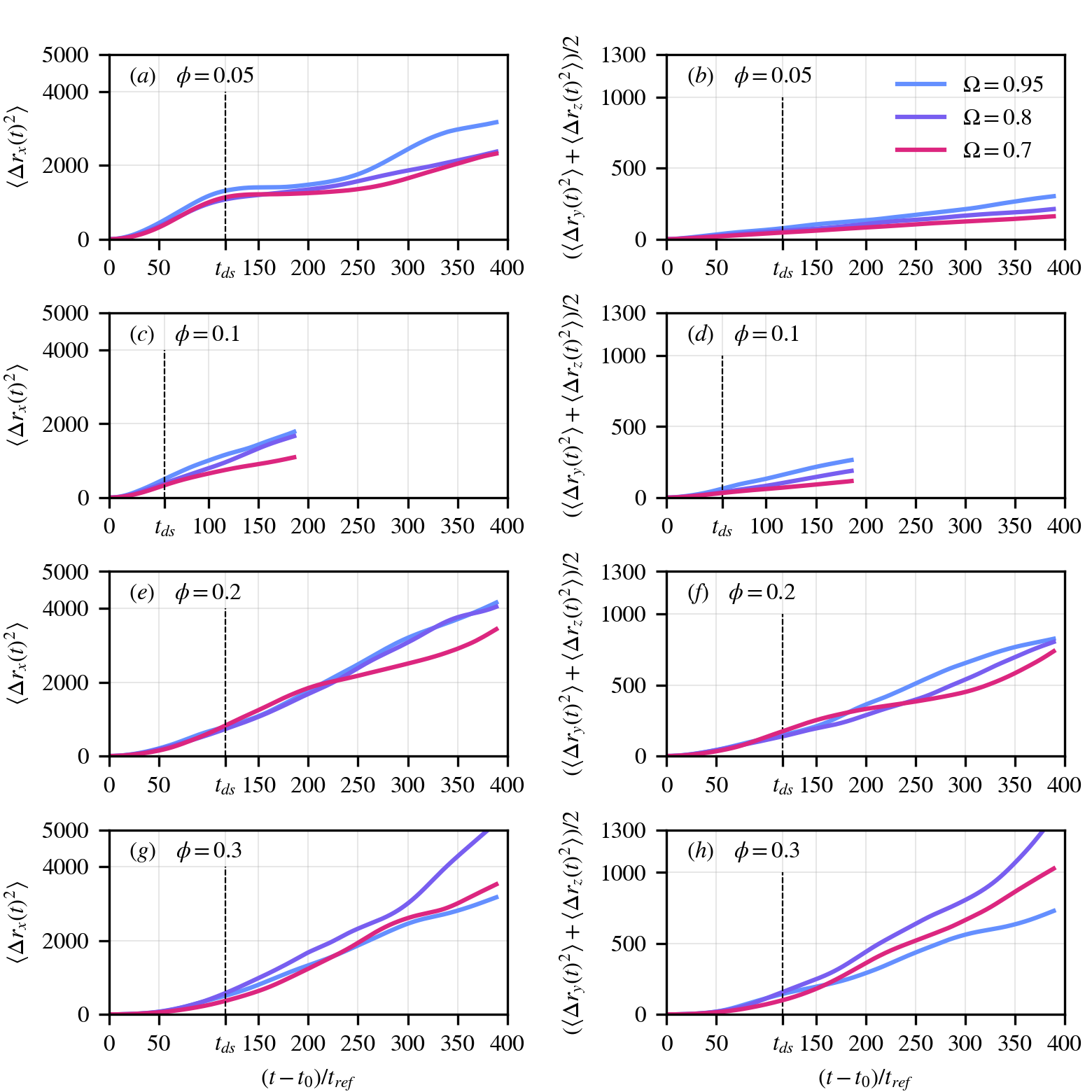}
    \caption{Mean square displacements in vertical (left panels) and in horizontal (right panels) directions for $\Omega \in \{ 0.95,0.8,0.7\}$ and $\phi \in \{0.05,0.1,0.2,0.3 \}$. \am{Dashed vertical lines indicate the value of time $t_{ds}$ that is used for the calculation of self-diffusion coefficients.}
    }
    \label{fig:msd_plot}
\end{figure}

Particle dispersion provides a quantitative measure of the long-time stochastic motion of particles induced by hydrodynamic interactions and collective effects in a settling suspension. In particular, the self-diffusion coefficient characterizes the irreversible spreading of particle trajectories about their mean settling motion and serves as a key descriptor of mixing and transport properties in particulate flows. We computed the self-diffusion coefficient tensor $\mathbf{D}$ of porous particles from their mean-square displacements (MSD), following the approach of \citet{yin2008velocity}:
\begin{equation}
\label{eq:self_diffusion}
\begin{aligned}
\mathbf{D} &=
\lim_{t \to \infty} \frac{\langle \Delta \mathbf{r}(t)^2\rangle}{2t} = \lim_{t \to \infty} \frac{1}{2t}\,\big\langle \big[\mathbf{X}(t) - \mathbf{X}(t_0) - \langle \mathbf{U} \rangle t \big] \otimes \big[\mathbf{X}(t) - \mathbf{X}(t_0) - \langle \mathbf{U} \rangle t \big] \big\rangle ,
\end{aligned}
\end{equation}

%
%\bv{I know we talked about this, but we may not need the full expansion of the tensor notation of $D_{ii}$. This would make equation 3.4 more compact. What do you think?} \am{Actualy I like it like this, to be maximal explicit, but we can shorten it as well, for example eliminating term 4}
where $\langle \Delta \mathbf{r}(t)^2 \rangle$ is the ensemble-averaged mean-square displacement tensor, $\langle \Delta \mathbf{r}(t) \rangle$ denotes the ensemble-averaged displacement vector, 
%\bv{BV: How is this a tensor? Do you have a reference for this equation?}
%\am{Modified equation \eqref{eq:self_diffusion}, to explicitly show how it is a tensor by adding a tensor product $\otimes$ in equation.}
$\mathbf{X}(t)$ is the particle position vector at time $t$, 
%\bv{BV: what do you mean by "unwrapped"? Found the definition further below, but I think it can be skipped because it makes things needlessly complicated.}
and $\mathbf{X}(0)$ is the vector of the initial particle position, taken at $t_0 = 17t_{\text{ref}}$. 
%Unwrapping of particle positions reconstructs continuous trajectories by correcting the artificial jumps that occur when particles cross periodic boundaries, ensuring that displacements represent the true motion. We denote the unwrapping operator with a hat, i.e. $\hat{\cdot}$.
%\sout{Since our data set is finite in time, we calculate the self-diffusion coefficient over the interval $[t_{ds}, t_{end}]$, where $t_{ds} = 68t_{ref}$ is chosen to exclude the initial transient after particle initialization, during which the mean-square displacement remains in the ballistic regime, so that diffusion statistics are evaluated only in the well-developed diffusive regime.}
%\bv{BV: explain why you start at 68.}
Although diffusion is formally described by a second-order tensor, in the present study, we focus only on its vertical and horizontal components, which correspond to the main diagonal of the diffusion tensor. Owing to the presence of gravity, particle settling occurs along the $x$-axis, and the dynamics in this direction are therefore expected to differ from those in the transverse directions. In contrast, the $y$- and $z$-directions are physically equivalent, because the simulation domain utilizes triple periodic BC, and there is no mechanism that breaks isotropy in the plane perpendicular to gravity. Hence, the horizontal diffusion can be characterized by a single effective coefficient obtained by combining the $y$- and $z$-components of the diffusion tensor, leading to the definitions $D_\parallel = D_{xx}$ and $D_\perp = (D_{yy} + D_{zz})/2$. 
\am{We calculate the self-diffusion coefficient over the interval $[t_{ds}, t_{end}]$, where $t_{ds} = 0.3 (t_{end}-t_0)$ is chosen to exclude the initial transient after particle initialization, during which the mean-square displacement remains in the ballistic regime. The value of $t_{ds}$ is shown as a vertical dashed line for each panel in figure \ref{fig:msd_plot}. The vertical and horizontal self-diffusion coefficients are obtained as the slope of a least-squares linear fit of the target functions:
\begin{eqnarray}
    f_{t,\parallel} (t) &=& 2 D_\parallel \langle \Delta r_x(t)^2 \rangle + const., \label{eq:diff_linear_fit_func} \\ 
    f_{t,\perp} (t) &=& 2 D_\perp \left( \frac{\langle \Delta r_y(t)^2 \rangle + \langle \Delta r_z(t)^2 \rangle}{2} \right) + const., \label{eq:diff_linear_fit_func_2}
\end{eqnarray}
over the time interval $t \in [t_{ds},t_{end}]$. The constants in \eqref{eq:diff_linear_fit_func} and \eqref{eq:diff_linear_fit_func_2} arise because the linear fit functions $f_{t,\parallel} (t)$ and $f_{t,\perp} (t)$ are evaluated from $t_{ds}$ rather than $t_0$, and therefore are not constrained to vanish at zero displacement. However, the values of these constants are not of interest for the further analysis below, as they do not affect the estimated self-diffusion coefficients.}
%\am{Explain why there is a constant and why it is not equal to zero. Link it to the ballistic regime.}
%The horizontal diffusivity $D_{\perp}$ is defined analogously from the target function $f_{t,\perp}(t) = \big( \langle \Delta r_y(t)^2 \rangle + \langle \Delta r_z(t)^2 \rangle \big) / 4$.
%\am{$f_{\parallel}$ should be given in form ax+b. Then, the plot with delta r need to be expressed in terms of this function in order to avoid confusion with denominator (2 for parallel and 4 for horizontal)}

%\bv{BV: you only take values at discrete times? I thought one has to integrate $\Delta r$ in time.}
%\am{AM: the original relation \ref{eq:self_diffusion} does not suggest integration.}

\begin{figure}
    \centering
    \includegraphics[]{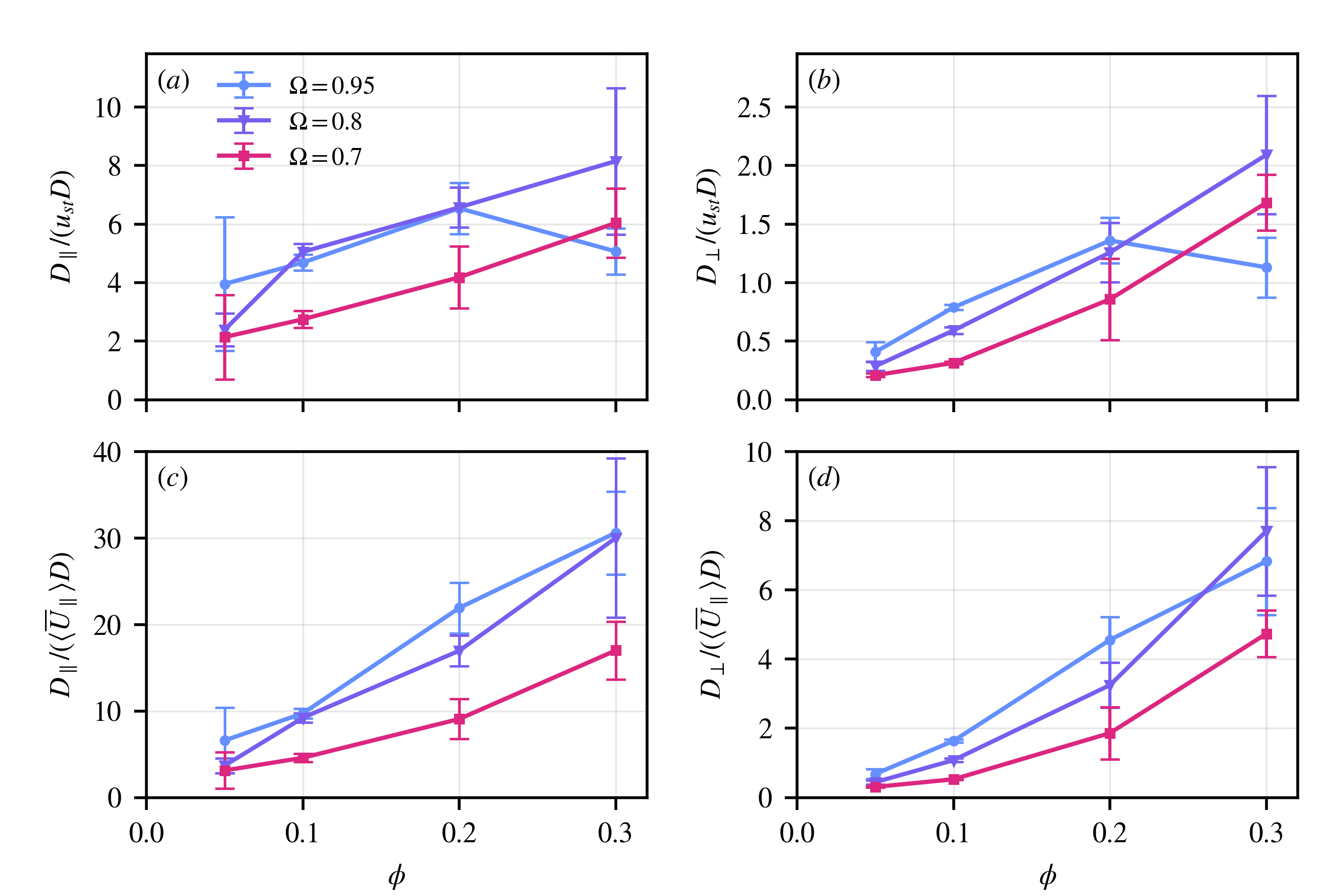}
    \caption{Vertical $(a,c)$ and horizontal $(b,d)$ components of particle self diffusion coefficient normalized by the constant reference velocity $(a,b)$ and the settling velocity of suspension $(c,d)$.
    }
    \label{fig:diff_coef}
\end{figure}

The MSDs for the vertical and horizontal directions are presented in figure \ref{fig:msd_plot} for volume fractions $\phi \in \{0.05, 0.1, 0.2, 0.3\}$ and drag reduction factors $\Omega \in \{0.95, 0.8, 0.7\}$.  
%\bv{BV: yet another way of denoting directions...}.
In all cases, the MSDs increase monotonically with time, indicating a persistent dispersive motion of particles in the vertical and horizontal directions. 
%\bv{BV: this does not come by surprise if you keep adding quadratic terms.}
As expected, the vertical MSDs are systematically larger than their perpendicular counterparts, reflecting the anisotropy induced by gravity. Moreover, the horizontal MSDs are more sensitive to changes in particle volume fraction, showing a markedly steeper increase at $\phi = 0.3$ compared with $\phi = 0.05$ for all values of $\Omega$. As shown in figure \ref{fig:msd_plot}, for all parameter combinations the MSD exhibits an overall linear increase in time, with superimposed fluctuations over the time interval $t_{ds} \le t \le t_{end}$. These fluctuations arise from the finite size of the computational domain and are expected to diminish as the number of particles and the total simulation time approach infinity. In practice, however, such limits are unattainable, and therefore quantifying these fluctuations is necessary to estimate the uncertainty of the diffusion coefficient. To assess this uncertainty, we evaluate the time derivative of the MSD at each Lagrangian output time step within $t_{ds} \le t \le t_{end}$. The standard deviation of these derivative samples is then computed and used as a measure of the fluctuation amplitude, which is shown as error bars in figure \ref{fig:diff_coef}.

Figure \ref{fig:diff_coef} shows the dependence of the diffusion coefficient components, calculated via \eqref{eq:self_diffusion}, on particle volume fraction. The panels $(a)$ and $(c)$ display the vertical component of diffusion coefficient $D_\parallel$, while the panels $(b)$ and $(d)$ correspond to the horizontal component $D_\perp$. In panels $(a)$ and $(b)$, the coefficients are normalized by a constant reference settling velocity $u_{st}$, and in panels $(c)$ and $(d)$ the values are normalized by the converged settling velocity specific to each simulation case. From panels $(a)$ and $(b)$, both diffusion components exhibit distinct trends. For the most impermeable particles ($\Omega = 0.95$), the diffusion coefficient increases with $\phi$, reaching a maximum at $\phi = 0.2$ before decreasing at $\phi = 0.3$. 
For more permeable particles, a similar pattern is observed; however, the maximum shifts to higher particle volume fractions, with the peak occurring at $\phi = 0.3$. It is expected that at higher $\phi$ values the diffusion coefficient would decline, as particle mobility diminishes with increasing solid volume fraction. Notably, the volume fraction corresponding to the maximum diffusion differs depending on particle permeability. Another important observation is that, at low particle volume fractions, more impermeable particles tend to exhibit higher diffusion coefficients. Specifically, for $0.05 < \phi < 0.2$, figure \ref{fig:diff_coef}$(a)$ shows nearly identical $D_\parallel$ values for $\Omega = 0.95$ and $\Omega = 0.8$, whereas the corresponding horizontal component is higher for $\Omega = 0.95$ over the same range. For all conditions, the case with $\Omega = 0.7$ consistently exhibits the lowest values of both components of the tensor $\bm{D}$. 

When the diffusion coefficients are normalized by the case-specific settling velocity, i.e. $\overline{U}_\parallel$ in panels \ref{fig:diff_coef}$(c)$ and $(d)$, a monotonic increase with $\phi$ is observed in all cases. 
Notably, for $\Omega = 0.95$ the temporal evolution of both diffusion coefficients remains approximately linear and does not exhibit a reduction at $\phi = 0.3$. This behaviour can be attributed to the fact that the decrease in the mean settling velocity from $\phi = 0.2$ to $\phi = 0.3$ (see figure \ref{fig:settl_vel_over_time}) is more pronounced than the corresponding decrease in the diffusion coefficients. Under this normalization, the maximum values for $\Omega = 0.95$ shift to $\phi = 0.3$, where they become comparable to those for $\Omega = 0.8$.

In summary, the analysis reveals that particle self-diffusion is anisotropic, with vertical displacements systematically exceeding horizontal ones, while generally increasing with particle volume fraction. Crucially, increased particle permeability suppresses this diffusive motion, resulting in consistently lower self-diffusion coefficients for the most permeable particles compared to the nearly impermeable cases across all investigated volume fractions.
\am{However, for $\phi = 0.3$ both components of self-diffusion coefficient are smaller for $\Omega = 0.95$ compared to the more permeable case $\Omega =0.8$. We conclude that this behaviour can be attributed to two competing mechanisms. First, cases with lower permeability exhibit more energetic relative counterflows, as shown in figure \ref{fig:counter_flows}$b$, which provide a greater potential to displace particles both opposite to the settling direction and in the horizontal directions. Although the intensity of the relative counterflows increases with increasing $\Omega$, the overall energy dissipation within the suspension also becomes larger. This dissipation arises from hydrodynamic and contact interactions between particles, where the latter becomes dominant for suspensions of $\phi \geq 0.3$ \citep{vowinckel2021rheology}. Consequently, a larger fraction of the flow energy is dissipated, leaving less energy available for particle motions that hinder settling, thereby counteracting the influence of the relative counterflows.}

\subsection{Suspension microstructure}
To examine how particles organize during settling, we conduct two complementary analyses. First, we perform a Voronoï tessellation based on particle positions to evaluate the distribution of Voronoï cell volumes, providing insight into the global particle arrangement and clustering within the domain. Second, we compute particle–particle distribution maps to characterize the local structural organization.
\subsubsection{Voronoï tesselation}
\begin{figure}
    \centering
    \includegraphics[]{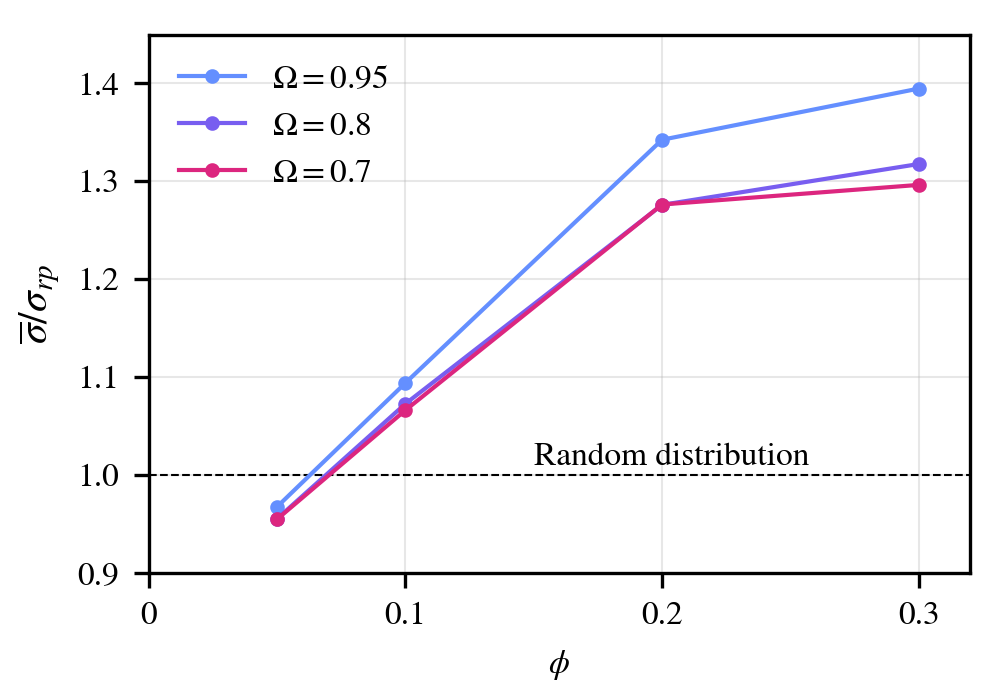}
    \caption{The standard deviation of the Voronoï cell volumes distribution averaged over time for various values of $\Omega$ and $\phi$.
    %\bv{BV: should be $\overline{\sigma}$}
    %\bv{BV: Mention for (c) that the red and purple line collapse}
    }
    \label{fig:voronoi_tessellation}
\end{figure}
Within this analysis, the domain is partitioned into cells such that the centroid of each cell coincides with the position of a particle center. By construction, every point within a given cell is closer to its corresponding centroid than to any other. A key advantage of this approach is that it does not require the specification of an \emph{a priori} characteristic length scale, thereby reducing ambiguity in the analysis. Voronoï tessellation is performed at each Lagrangian output time step. 
%\bv{BV: I assume again that you have regular intervals. State the outputting interval.}
The volume of each cell is then normalized by the average volume of the cell. From the resulting distribution of normalized cell volumes, we compute the standard deviation $\sigma$, which quantifies the spread of local cell volumes relative to the mean.
The quantity $\sigma$ provides a compact statistical measure of spatial heterogeneity in the particle distribution, with larger values indicating stronger clustering.
The time-averaged value of $\overline{\sigma}$ is evaluated over the interval $150 < \tilde{t} < 200$ for simulations with $\phi = 0.1$, and over $350 < \tilde{t} < 400$ for simulations with $\phi \in \{0.05,0.2, 0.3\}$. These averaging intervals correspond to the statistically steady state, in which the mean value of $\sigma$ remains constant over time.
The standard deviation calculated for the initial particle positions and denoted as $\sigma_{rp}$ corresponds to the value of the particles that are distributed using the random sequential addition algorithm.
We consider the value of $\sigma_{rp}$ as the reference standard deviation corresponding to a random distribution of spheres. The values of $\sigma_{rp}$ are listed in table \ref{tab:sigma_rp}. They depend only on $\phi$ and are identical for all values of $\Omega$, since the initial particle distribution is unchanged across different $\Omega$. Values of $\overline{\sigma}$ exceeding $\sigma_{rp}$ indicate the emergence of clustering, as they reflect an increased occurrence of both densely populated regions (small Voronoï volumes) and voids (large Voronoï volumes). Conversely, values of $\overline{\sigma}$ smaller than $\sigma_{rp}$ imply a tendency toward a more uniform or ordered particle arrangement, characterized by reduced spatial clustering.

\begin{table}
    \centering
    \begin{tabular}{c|cccc}
        $\phi$ & $0.05$ & $0.1$ & $0.2$ & $0.3$ \\
        \hline
        $\sigma_{rp}$ & $0.334$ & $0.27$ & $0.176$ & $0.128$ \\

    \end{tabular}
    \caption{Standard deviations of Voronoï cell volumes computed using the initial position of particles for various particle volume fractions $\phi$.
    %\bv{BV: I think the horizontal line is needed. }
    }
    \label{tab:sigma_rp}
\end{table}

The ratio of $\overline{\sigma}/\sigma_{rp}$ is shown in figure \ref{fig:voronoi_tessellation} as a function of volume fraction $\phi$ for all simulation cases with $\phi \ge 0.05$ in table \ref{tab:simulation_parameters}. 
It is visible from figure \ref{fig:voronoi_tessellation} that the ratio of $\overline{\sigma}/\sigma_{rp}$ increases with higher particle volume fraction for all values of $\Omega$. In addition, it is also seen that the values of $\overline{\sigma}$ increase with higher $\Omega$.
The close similarity between the cases $\Omega = 0.8$ and $\Omega = 0.7$ persists across the entire range of $\phi$. This indicates similarity in microstructure between the more permeable particle properties.
Overall, these results show that increasing particle volume fraction promotes the development of clustered microstructures, as reflected by the increasing values of $\overline{\sigma}/\sigma_{rp}$. In contrast, increasing permeability (lower $\Omega$)
%\bv{BV: I thought decreasing $\Omega$ increases permeability?}
counteracts this clustering tendency, leading to more homogeneous particle arrangements. 
Decreasing $\Omega$ from 0.8 to 0.7 does not further amplify the effect.

\subsubsection{Pairwise particle distribution}
\begin{figure}
    \centering
    \includegraphics{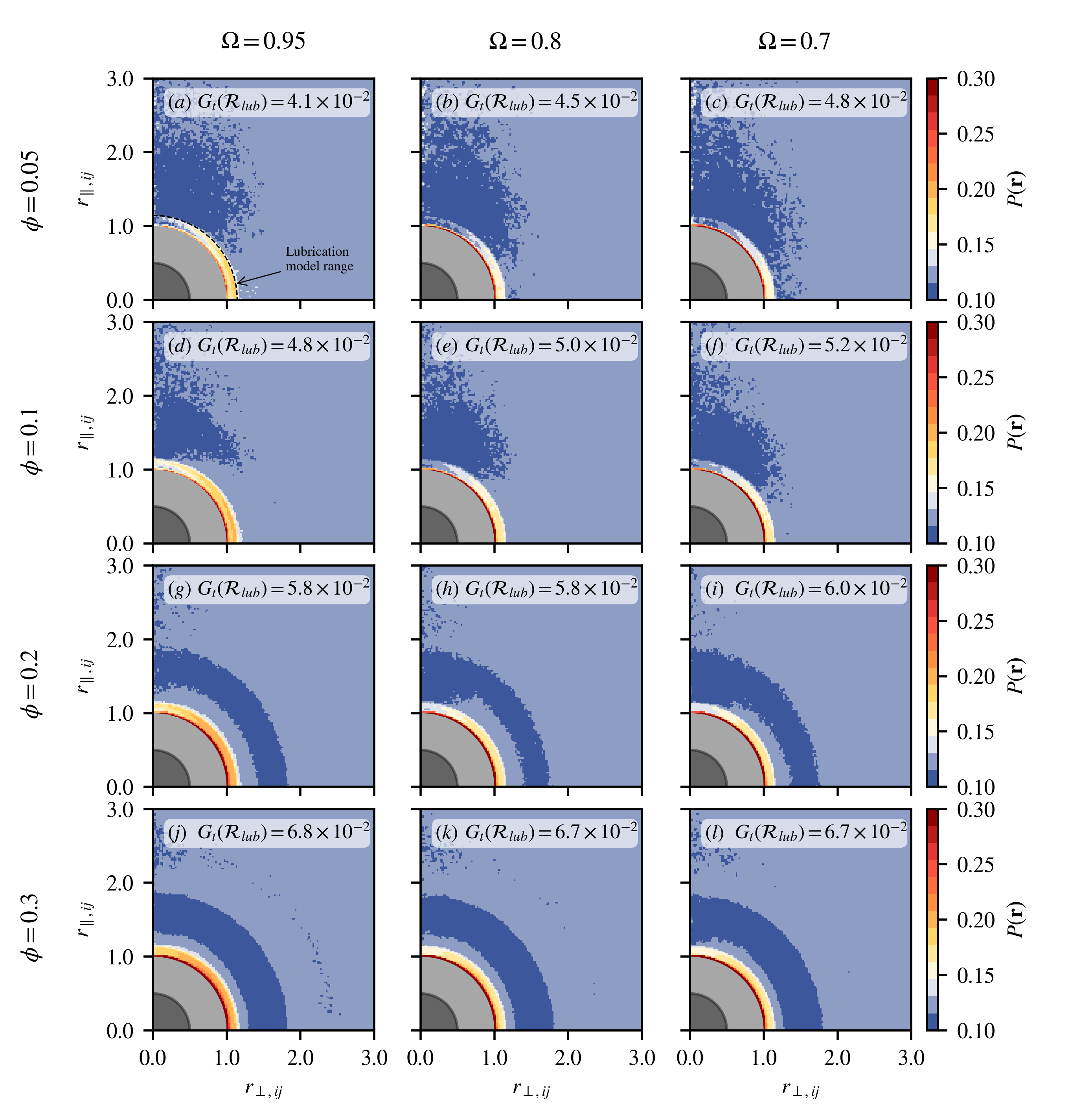}
    \caption{Pairwise particle distribution $P(\bm{r}_{ij})$ as a function of $r_\parallel$ and $r_\perp$ coordinates.
    %\bv{BV: I see all cases up to $\phi=0.3$}
    The results are presented for all values of $\Omega$ and $\phi$. The dark gray area represents the reference particle, while the light gray area represents the area of no collision.
    The probability of finding a neighbouring particle in the lubrication range, $G_t(\mathcal{R}_{lub})$ according to \eqref{eq:lb_range_neighbor}, is additionally given as numerical values in each panel.}
    \label{fig:pairwise_distr_all_phi}
\end{figure}

Examining particle distributions on the local scale for different values of $\phi$ and $\Omega$ may help to understand the differences in $\overline{\sigma}/\sigma_{rp}$ shown in figure \ref{fig:voronoi_tessellation} for the analysis of the Voronoï tesselation. 
%\bv{BV: do you mean panel (a) and (b) or is this more about panel (c)? In the latter case, calling it $\sigma/\sigma_{rp}$ would be more precise.}
Towards this end, we plot the pairwise particle density distribution function $P_b(\bm{r}_{ij})$, defined in terms of the relative distance vector $\bm{r}_{ij}$ between particle pairs $i$ and $j$ in figure \ref{fig:pairwise_distr_all_phi} as a means to analyze the suspension microstructure \citep[c.f.][]{yin2007hindered,uhlmann2014sedimentation,shajahan2020influence}. The function is evaluated on a two-dimensional uniform rectangular grid, where the subscript $b$ denotes a specific grid cell. The distance vector $\bm{r}_{ij}$ is expressed in non-dimensional coordinates $(r_{\perp,ij}, r_{\parallel,ij})$. 
%\bv{BV: is there a reference we can provide for this analysis tool?}
The relative distance between the particles $i$ and $j$ is defined as 
\begin{equation}
    \label{eq:relative_position}
   \bm{r}_{ij} = (r_{\perp,ij}, r_{\parallel,ij})^T = (\sqrt{\tilde{Y}_i^{2}+\tilde{Z}_i^{2}} - \sqrt{\tilde{Y}_j^{2}+\tilde{Z}_j^{2}}, \, \tilde{X}_i - \tilde{X}_j)^T \quad, 
\end{equation}
where $\tilde{X}$, $\tilde{Y}$, and $\tilde{Z}$ denote the non-dimensional coordinates of the particle centers.
%\bv{Here, you introduce indices as you should have done it throughout your manuscript}
%\am{now addressed}
Similarly to the diffusion coefficient, the relative distances are analyzed component-wise, distinguishing the vertical ($x$) direction of settling. The horizontal components ($y$ and $z$) are combined to reflect the radial distance from the settling axis. This treatment is justified by the axisymmetry of the simulation domain around the settling direction, which makes $y$ and $z$ equivalent and eliminates any preferential direction in the plane perpendicular to $x$. 
%\am{The indices $i,j$ are omitted for the remainder of the present section for notational simplicity.}
%\bv{Please keep them}
The pairwise particle density distribution function is computed over the region $\mathcal{R}_{dom} = \{ (r_{\perp,ij}, r_{\parallel,ij}) \mid 0 < r_{\perp,ij} < 3, 0< r_{\parallel,ij} < 3 \}$, and is given by
\begin{equation}
\label{eq:pairwise_distribution}
P_b(\bm{r}_{ij}) = \frac{1}{\Delta b^2 N_t n_p(n_p-1)} \sum_{k=1}^{N_t} \sum_{i=1}^{n_p} \sum_{\substack{ j \neq i}}^{n_p} K\left(\frac{||\bm{r}^b_{ij} - \bm{r}_{ij}||_\infty}{\Delta b}\right) \quad,
\end{equation}
such that the total probability $G_t(\mathcal{R}_{dom})$ of finding a neighbor satisfies
\begin{equation}
    \label{eq:total_probability}
    G_t(\mathcal{R}_{dom}) = \int_{\mathcal{R}_{dom}} P_b(\bm{r}_{ij}) \, d\bm{r}_{ij} = 1 \quad .
\end{equation}
In equation \eqref{eq:pairwise_distribution} $\Delta b = 1/50$ is the spatial discretization of $P_b$ and $\bm{r}^b_{ij}$ denotes the center of the auxiliary grid cell used for the discretization of $P_b(\bm{r}_{ij})$. The kernel function $K(q)$ in \eqref{eq:pairwise_distribution} is defined as
\begin{equation}
\label{eq:kernel}
K(q) =
\begin{cases}
1, & \text{if } |q| < 1/2, \\
0, & \text{otherwise}.
\end{cases}
\end{equation}
Here, $q$ denotes a generic non-dimensional argument of the kernel. For reference, a perfectly uniform particle distribution corresponds to a constant probability density, $P_b(\bm{r}_{ij}) = 1/(9-\pi/4) \approx 0.12$, implying an equal likelihood of finding a neighboring particle at any relative position $\bm{r}_{ij}$. 
Note that, due to the triple periodic boundary conditions, the closest particle position $\bm{r}_{ij}$ relative to the bin position $\bm{r}_{ij}^b$ is used when evaluating the distance $||\bm{r}_{ij}^b - \bm{r}_{ij}||_\infty$ in equation \eqref{eq:pairwise_distribution}.

It is seen from figure \ref{fig:pairwise_distr_all_phi} that the highest values of probability to find a neighboring particle around the reference particle are in the close proximity of the reference particle. This proximity correlates well with the range of the lubrication model, which acts within the spherical shell of two grid cell thickness $2\Delta h$ surrounding every particle.
The external boundary of the lubrication model between two particles is shown as a dashed arc on figure \ref{fig:pairwise_distr_all_phi}$(a)$. 
We also introduce an additional metric to estimate the total probability of finding a neighboring particle within the lubrication model range as 
\begin{equation}
    \label{eq:lb_range_neighbor}
    G_t(\mathcal{R}_{lub}) = \int_{\mathcal{R}_{lub}} P_b(\bm{r}_{ij}) \, d\bm{r}_{ij} \quad,
\end{equation}
where $\mathcal{R}_{lub}$ is the area of relative positions where the lubrication model between the reference particle and the neighboring particle is modeled, i.e. $\mathcal{R}_{lub} = \{ (r_\perp, r_\parallel) \mid D < ||\bm{r}_{ij}||^2 < D+2 \Delta h \}$.

The probability of finding a neighboring particle within the range of the lubrication model remains consistently high across all simulation cases. Nevertheless, the spatial distribution of neighbors exhibits clear variations with respect to both $\Omega$ and $\phi$. For the simulations at $\phi = 0.05$ (figures \ref{fig:pairwise_distr_all_phi}$(a,b,c)$), the likelihood of encountering a neighbor aligned horizontally is greater than that of a neighbor aligned vertically. Furthermore, the likelihood of encountering a neighbor within the lubrication range increases with decreasing $\Omega$, as shown by the corresponding value of $\mathcal{R}_{lub}$ in figure \ref{fig:pairwise_distr_all_phi}. A similar trend is observed for $\phi = 0.1$, while for higher particle volume fractions $G_t(\mathcal{R}_{lub})$ remains nearly constant across all values of $\Omega$. 
%\am{On the other hand, similarly as in figure \ref{fig:voronoi_tessellation}, the notable difference is observed in the distribution of $P_b(\bm{r})$ between all simulations with $\Omega = 0.95$ and cases with $\Omega \in \{0.8,0.7\}$. The difference is observed in $P_b(\bm{r})$ distribution primarily in the lubrication model range. While for all simulation cases the peak of $P_b(\bm{r})$ is observed in the direct vicinity of the reference particle, that peak is weaker for $\Omega = 0.95$ in comparison with values of $\Omega \in \{0.8,0.7\}$. However, with increasing distance from the reference particle, $P_b(\bm{r})$ decreases faster for $\Omega \in \{0.8,0.7\}$. }
%\bv{BV: you can tie this observation to your argument regarding the Voronoii tesselation and compare it to your conclusion of more clustering for higher $\phi$ but a reduction for decreasing $\Omega$}
%\am{I didn't understand this comment.}

We attribute the observed increase of $G_t(\mathcal{R}_{lub})$ with decreasing $\Omega$ primarily to a modification of the classical drafting–kissing–tumbling (DKT) mechanism, which was originally described by \citet{fortes1987nonlinear}. 
In DKT settling for impermeable spheres, a trailing particle enters the wake of a leading one, accelerates due to the reduced hydrodynamic resistance (“drafting”), makes near contact (“kissing”), and subsequently separates (“tumbling”). The key stage controlling the pair separation is the tumbling phase, during which the particles orient approximately horizontally relative to the settling direction. In this configuration, the pressure build-up in the narrow gap between them generates a repulsive hydrodynamic force that drives them apart. For more permeable particles (lower $\Omega$), the interstitial flow within and around the particles mitigates the local pressure rise in the gap, thereby weakening this repulsive force. As a consequence, the tumbling-induced separation becomes less effective, resulting in a more sustained hydrodynamic interaction between neighboring particles. This finding suggests that the DKT sequence might be strongly affected by particle permeability.

A second trend emerges when comparing $P_b(\bm{r}_{ij})$ qualitatively within the lubrication model range for simulations with different values of $\Omega$. Although the maxima of the pairwise particle density distribution $P_b(\bm{r}_{ij})$ remain located near the contact line in all cases, the rate at which they decay with increasing $\bm{r}_{ij}$ depends strongly on particle permeability. In particular, for $\Omega \in \{0.7,0.8\}$ the maxima decay more rapidly away from the contact line than for $\Omega = 0.95$ (the contact line corresponds to the outer boundary of the light gray quarter circle). The similarity between cases with $\Omega \in \{0.7,0.8\}$ is consistent with the trends observed in the Voronoï tessellation volume analysis shown in figure \ref{fig:voronoi_tessellation}, where the standard deviation $\overline{\sigma}$ of simulations with $\Omega \in \{0.7,0.8\}$ are almost identical to each other. This suggests that the exact distribution of $P_b(\bm{r}_{ij})$ in the immediate vicinity of the reference particle may govern the resulting distribution of Voronoï volumes. At the same time, decreasing $\Omega$ leads to an increase in the maximal values of $P_b(\bm{r}_{ij})$ along the contact line. We attribute the difference in $P_b(\bm{r}_{ij})$ distribution in the vicinity of the reference particle to the lubrication interactions between neighbouring particles. The lubrication force scales inversely with the interparticle gap, $\bm{F}_{lub} \propto 1/\zeta$, and attains its maximum at $\zeta_n^p$. For more permeable particles, the characteristic cutoff distance $\zeta_n^p$ is larger, resulting in weaker lubrication interactions. Conversely, less permeable particles experience stronger lubrication forces during close particle-particle interactions, leading to enhanced dissipation of relative momentum. This suppresses particle separation after near-contact interactions and therefore increases the probability of finding neighbouring particles at small interparticle distances.

Overall, these results demonstrate that particle permeability modifies both the strength and spatial persistence of near-field hydrodynamic interactions, thereby altering the classical DKT dynamics and directly shaping the microstructural organization of particle pairs across different volume fractions.

\section{Conclusion and outlook}
\label{sec:conclusion}

The current study presents results of direct numerical simulations of suspensions settling under gravity. The suspension consists of highly porous particles to investigate the combined influence of particle permeability ($\Omega$) and particle volume fraction ($\phi$) on the settling dynamics and the related changes in the suspension microstructure.

The results established that the relationship between particle volume fraction $\phi$ and suspension settling velocity $\langle U_\parallel \rangle$ is highly affected by particle permeability. As volume fraction increases, differences in settling velocity become more pronounced, reaching an approximate difference of up to $106\%$ between the least permeable particles ($\Omega = 0.95$) and the most permeable particles ($\Omega = 0.7$), as investigated here at $\phi = 0.3$. The observed hindered settling behavior in the tested range ($0.05 < \phi < 0.3$) is well described by the Richardson-Zaki equation \eqref{eq:RZ_equation}. The fitted RZ exponent ($n_{rz}$) systematically decreased with decreasing $\Omega$, ranging from $n_{rz}=4.2$ for $\Omega=0.95$ to $n_{rz}=2.15$ for $\Omega=0.7$. This dependence was further generalized by a power-law function $n_{rz} = a_f\Omega^{b_f} + 1$, with $a_f=3.82$ and $b_f=3.23$, fitted for $\Rey_{\parallel,0} = 0.85$. We further assume that this relation remains valid at lower $\Rey_{\parallel,0}$, since $n_{rz}$ converges to a constant value as $\Rey_{\parallel,0} \to 0$.

The study proposed that the decreased momentum exchange between counterflows and porous particles is the main mechanism responsible for the varying change in settling speed as a function of particle volume fraction. Quantitative analysis of the mean vertical fluid velocity, $u_{f, \parallel}$, confirmed this hypothesis. Suspensions composed of less permeable particles (higher $\Omega$) generate stronger counter flows with a higher relative upward velocity. When normalized by suspension settling velocity, the relative counterflow velocity differed by nearly a factor of two between the lowest ($\Omega=0.7$) and highest ($\Omega=0.95$) permeability cases at $\phi=0.3$. 

The velocity fluctuations, normalized by the settling of the suspensions, systematically increased with particle volume fraction in the range $0.05 < \phi < 0.3$, irrespective of permeability. Vertical and horizontal components of the velocity fluctuations reached their highest values for the least permeable particles ($\Omega = 0.95$) throughout the full range of particle volume fractions tested. Analysis of the self-diffusion coefficient showed that at low particle volume fractions ($0.05 < \phi < 0.2$), more impermeable particles exhibited higher diffusion coefficients compared to more permeable cases. The case with $\Omega = 0.7$ consistently exhibits the lowest values for both vertical and horizontal diffusion components under all conditions. When the diffusion coefficients were normalized by the case-specific settling velocity, a monotonic increase with particle volume fraction was observed in all cases.

The Voronoï tessellation analysis reveals that particles with the highest drag reduction factor ($\Omega=0.95$) exhibit a larger standard deviation of Voronoï cell volumes compared to cases where $\Omega \in \{0.7,0.8\}$. This indicates a more heterogeneous spatial arrangement at higher $\Omega$, which is consistent with the emergence of particle clustering, as an increased standard deviation reflects a higher probability of both densely populated regions and voids.
We explain this difference by analyzing the pair-distribution maps, which show that neighbors are more frequently located at shorter inter-particle distances for $\Omega=0.95$. However, for dilute suspensions ($\phi \in \{0.05,0.1\}$), the pair-distribution maps indicate that the probability of finding neighboring particles within the lubrication range increases as $\Omega$ decreases. We attribute this trend to changes in the drafting–kissing–tumbling (DKT) mechanism, which is modified by particle permeability.

%\backsection[Supplementary data]{\label{SupMat}Supplementary material and movies are available at \\https://doi.org/10.1017/jfm.2019...}

\backsection[Acknowledgements]{The authors thank Élisabeth Guazzelli for fruitful discussions that helped improve the manuscript. The authors gratefully acknowledge the computing time made available to them on the high-performance computer at the NHR Center of TU Dresden. This center is jointly supported by the Federal Ministry of Education and Research and the state governments participating in the NHR (www.nhr-verein.de/unsere-partner). The authors gratefully acknowledge the Gauss Centre for Supercomputing (GCS) e.V. (www.gauss-centre.eu) for funding this project by providing computing time on the GCS Supercomputer SUPERMUC-NG at Leibniz Supercomputing Centre (www.lrz.de).}

\backsection[Funding]{The current study is supported by the German Research Foundation (Grant No. VO2413/2-1).}

\backsection[Declaration of interests]{The authors report no conflict of interest.}

\backsection[Data availability statement]{
The data and the source code that support the findings of this study will be made available in the form of an open-source repository upon successful publication of the present study.
%The data that support the findings of this study are openly available in [repository name] at http://doi.org/[doi], reference number [reference number]. See JFM's \href{https://www.cambridge.org/core/journals/journal-of-fluid-mechanics/information/journal-policies/research-transparency}{research transparency policy} for more information
}

\backsection[Author ORCIDs]{A. Metelkin, https://orcid.org/0000-0001-6262-0924; B. Vowinckel, https://orcid.org/0000-0001-6853-7750}

\backsection[Author contributions]{A. Metelkin and B. Vowinckel conceived the study and defined the research objectives. A. Metelkin designed and implemented the numerical simulations, performed the analyses, and prepared the original draft of the manuscript. B. Vowinckel contributed to the interpretation of the results, revised the manuscript, and supervised the project.}

\appendix

\section{Validation of the simulation method}\label{appA}
\subsection{Settling of a particle}

The present validation test case involves the settling of a single particle in a quiescent fluid towards a wall. Experimental measurements of particle settling at various Reynolds numbers were reported by \cite{ten2002particle}. In their study, the authors provided the time evolution of the velocity of the particle from the moment of release to the point of contact with the bottom wall of the experimental setup. The experiments were conducted in a closed rectangular container of dimensions \(100\,\text{mm} \times 160\,\text{mm} \times 100\,\text{mm}\), with the height corresponding to the longest dimension. A spherical particle with a diameter of \(15\,\text{mm}\) was initially positioned at a height of \(120\,\text{mm}\) in the center of the domain and held in place using a Pasteur pipette via a negative pressure mechanism. Four different Reynolds numbers were tested in the experiments: \(\Rey_{\parallel,0} \in \{1.5,\,4.1,\,11.6,\,32.2\}\). Detailed descriptions of the properties of the fluid and particles are provided in \cite{ten2002particle}. In this study, we simulated two cases corresponding to \(\Rey_{\parallel,0} = 1.5\) and \(\Rey_{\parallel,0} = 4.1\), using a computational domain of the same size. The simulations were performed at two mesh resolutions: a fine grid with \(N_x \times N_y \times N_z = 200 \times 320 \times 200\), corresponding to 30 grid cells per particle diameter, and a coarse grid with \(94 \times 152 \times 94\), corresponding to approximately 14 grid cells per diameter. Since the original experiments used solid impermeable particles, the permeability of the simulated particles was adjusted to achieve a nearly impermeable condition, with \(\beta = 100\) and \(\Omega = 0.99\) for porosity $\epsilon = 0.95$. The effective density of the porous particle was also adjusted to match the density used in the experiments. 

\begin{figure}
    \centering
    \includegraphics[]{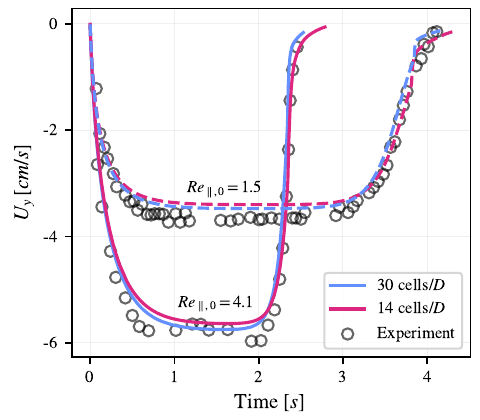}
    \caption{Settling velocity of a single particle in undisturbed flow. The simulation data from the present computational approach with different mesh resolution are compared to experimental data of \cite{ten2002particle} for $\Rey_{\parallel,0}=1.5$ and $\Rey_{\parallel,0}=4.1$}
    \label{fig:ten_Cate_comparison}
\end{figure}

Figure \ref{fig:ten_Cate_comparison} presents the comparison between the simulation results and the experimental measurements. For both Reynolds numbers considered, the simulation using the fine mesh more accurately captures the terminal settling velocity observed in the experiments. Nevertheless, the results of the coarse mesh also show good agreement, deviating by no more than 2.5\% in the terminal velocity. Overall, the comparison demonstrates good agreement between the numerical simulations and the experimental data, supporting the validity of the numerical approach employed.

\subsection{Flow field over a fixed particle}
We performed a qualitative analysis of the velocity patterns using simulation over a single fixed porous particle. We compared the velocity streamlines with the work of \cite{yu2012numerical}, where the authors analyze the flow around a single porous particle. We performed a simulation with the domain size of $5D\times5D\times 5D$ and placed the porous particle in the middle of the domain. The domain is discretized with $N_x \times N_y \times N_z = 150\times150\times150$ grid cells. The Reynolds number $\Rey_{in}$ for these simulations is based on the magnitude of inlet velocity $\bm{u}_{in}$ and defined as $\Rey_{in}=||\bm{u}_{in}|| D/\nu$. The system of governing equations for the present simulation was non-dimensionalized using $||\bm{u}_{in}||$, while the rest of the non-dimensional parameters are chosen similarly as in the main body of the paper. Thus, each non-dimensional quantity is defined consistently with \eqref{eq:Non_Dim_ref} and \eqref{eq:Non_Dim_ref_part}, with the only difference being that $u_{st}$ is replaced by $||\bm{u}_{in}||$.
%\bv{BV: refer to the exact equation}
The initial conditions were set to zero for the non-dimensional velocity and pressure, 
while the boundary condition on the left side of the computational domain was set to $\bm{u}_{in} = (u_{x,in},0,0)^T$ with $u_{x,in}$ chosen in the way to reach a specific value of $Re_{in}$. The boundary conditions for the outlet side of the domain are set as zero gradient, while for other surfaces of the domain, free-slip boundary conditions are applied. The boundary conditions for pressure were set as zero gradient for all surfaces in the computational domain. 
\begin{figure}
    \centering
    \includegraphics[]{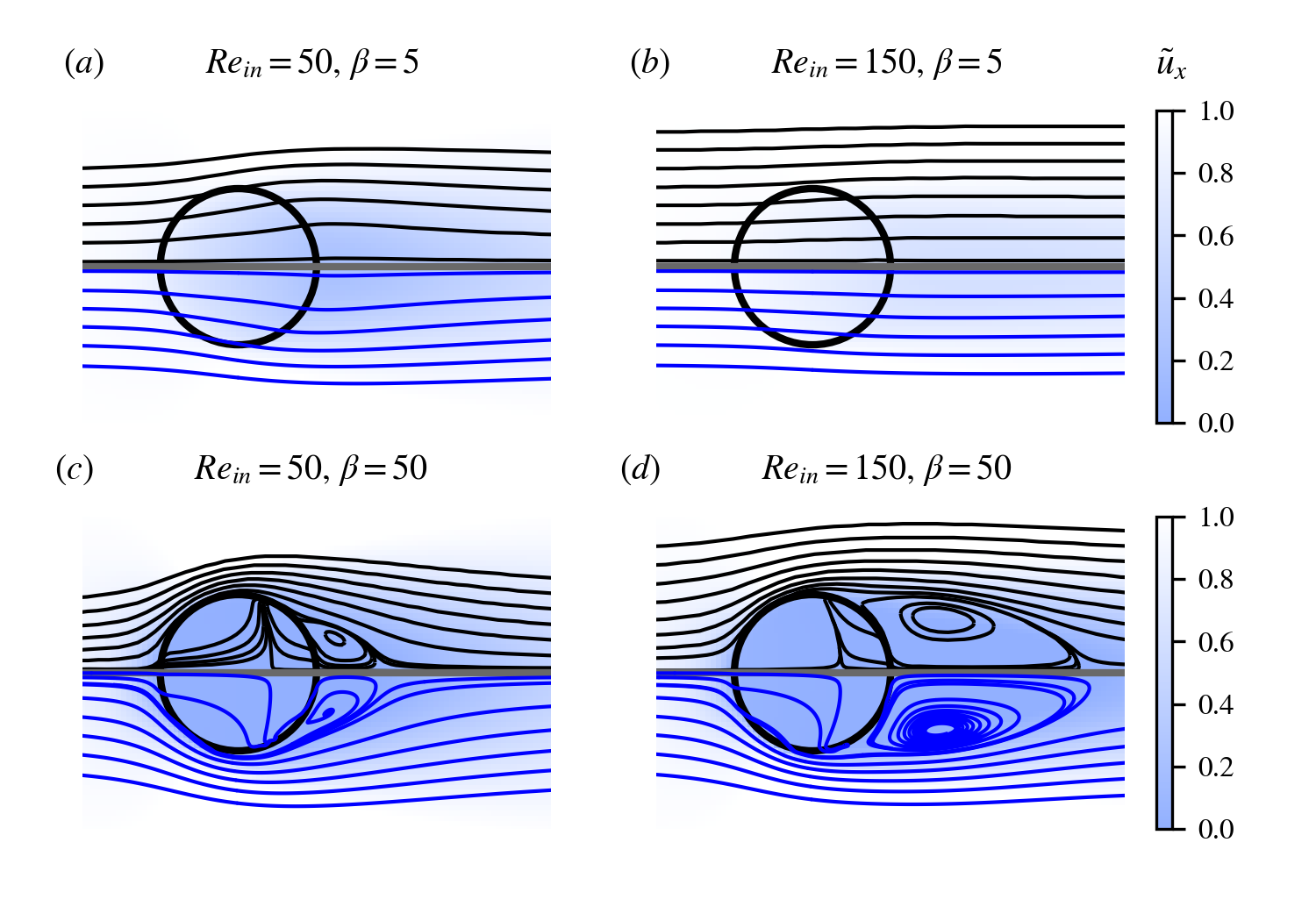}
    \caption{Comparison of the streamlines between the study of \cite{yu2012numerical} (black lines in the upper part) and the present simulation approach (blue lines in the lower part). The non-dimensional streamwise velocity component $\tilde{u}_x$ is shown as a contour plot}
    \label{fig:fixed_part_streamlines}
\end{figure}
Figure \ref{fig:fixed_part_streamlines} shows the comparison of the streamlines between the present simulations (blue streamlines on the lower part of each panel) and the streamlines extracted from the work of \cite{yu2012numerical} (black lines on the upper part of each panel). It is seen from figure \ref{fig:fixed_part_streamlines} that for all combinations of Reynolds number $\Rey$ and $\beta$, the streamlines of both simulations look very similar inside and outside of the porous particle. The simulation cases with $\beta = 50$ illustrate the same length of the recirculation zones behind the porous particle between the two studies. Both simulation scenarios characterized by higher permeability (i.e., $\beta = 5$) exhibit an almost negligible impact on fluid flow. Nonetheless, the minor alterations in the streamlines remain consistent between the studies. The close similarity of flow patterns additionally validates the simulation technique of the present study.

\section{Autocorrelation analysis}
\label{app:autocorrelation}
To choose the appropriate computational domain size, we performed the analysis based on the autocorrelation function of the Eulerian velocity for various domain sizes. The tested computational domains are listed in table \ref{tab:sim_dom_autocorr}.
\begin{table}
    \centering
    \begin{tabular}{c|ccc}
        Domain index & $L_x$ & $L_y$ & $L_z$ \\
        \hline
        $SD20$ & $20D$ & $20D$ & $20D$ \\
        $SD50$ & $50D$ & $30D$ & $30D$ \\
        $SD65$ & $65D$ & $65D$ & $65D$ \\
        $SD100$ & $100D$ & $50D$ & $50D$ \\
    \end{tabular}
    \caption{Simulation domains used for the autocorrelation analysis}
    \label{tab:sim_dom_autocorr}
\end{table}
The physical parameters of the fluid are the same as in the main body of the present study, and the physical parameters of particles correspond to those of $\Omega = 0.95$ from table \ref{tab:simulation_parameters}. The resolution of each grid is the same as for the main simulations of the present study, i.e., 14 grid cells per particle diameter. We set the constant value of the particle fraction for all simulation domains as $\phi = 0.05$. This choice is motivated by the study of \cite{shajahan2024interface}, where the authors demonstrate that the integral time scale of an autocorrelation function reaches the highest values for the lowest values of $\phi$. The autocorrelation based on Eulerian velocity $\bm{u}$ is denoted by $R_{u_i u_i}^e$ and is defined as follows:

\begin{equation}
    \label{eq:autocor_euler}
    R_{u_i u_i}^e (\Delta x_j) = \frac{1}{N_tN_{pl}} \sum_m^{N_t} \left[ \sum_k^{N_{pl}} \frac{ u_i'(x_{j,k}) u_i'(x_{j,k} + \Delta x_j)}{u_i'(x_{j,k})^2} \right]_m.
\end{equation}

%R_{u_i u_i}^e (r_j) = \sum_k^{\mathcal{L}} \frac{\langle u_i'(x_{j,k}) u_i'(x_{j,k} + r_j) \rangle}{\langle u_i'(x_{j,k}) u_i'(x_{j,k}) \rangle}.

\begin{figure}
    \centering
    \includegraphics[]{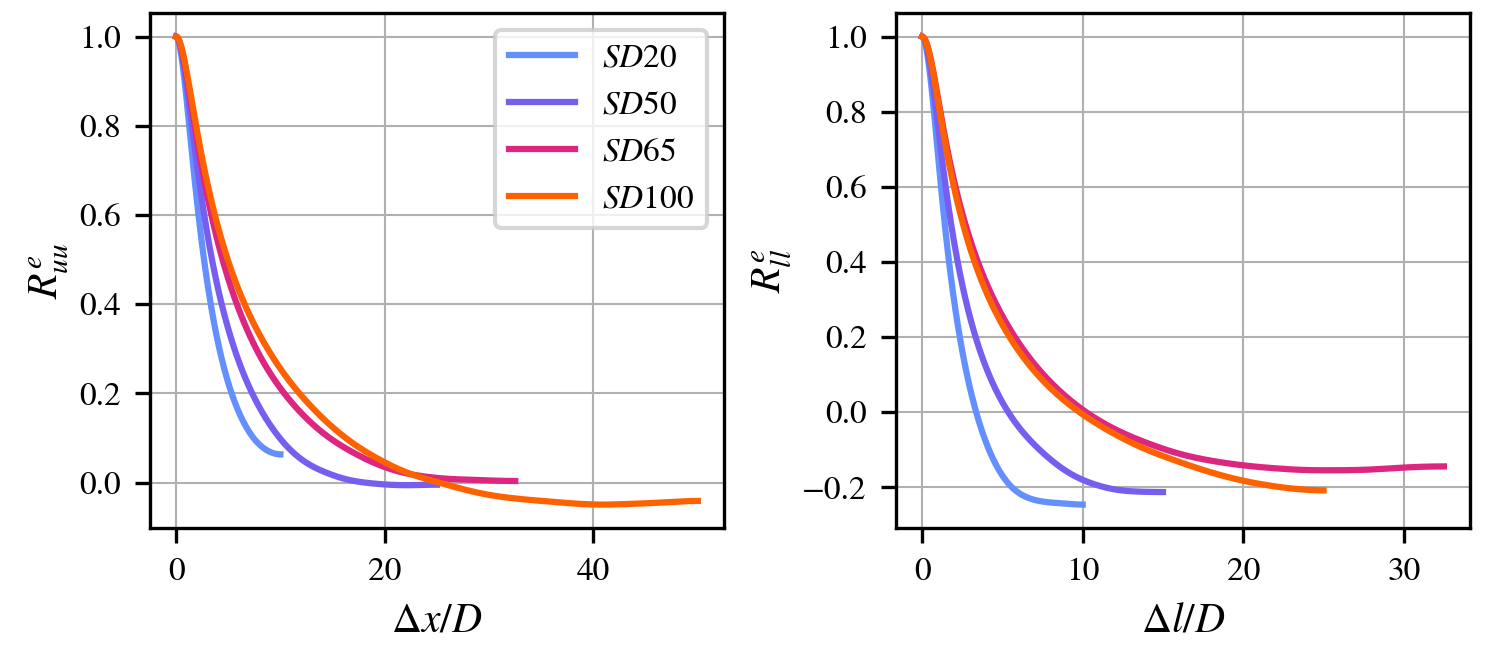}
    \caption{Autocorrelation functions $R_{uu}^e$ and $R_{ll}^e$ performed on simulations with $\phi = 0.05$ and $\Rey_{st} = 1$ as functions of vertical and horizontal components, respectively.}
    \label{fig:autocor_phi005}
\end{figure}

Here, $x_{j,k}$ represents a probe line whose direction is defined by the index $j$. The index $k$ refers to the horizontal position of the probe line. Probe lines were placed over the entire domain with a gap between each other of 10 grid cell sizes. The variable $u_i^\prime$ is the velocity fluctuation component computed for a specific time snapshot. The variables $N_t$ and $N_{pl}$ are the total number of time snapshots and probe lines, respectively. 
%\bv{BV: previously, $\mathcal{N}$ was called $N_t$.}
The variable $\Delta x_j$ corresponds to the space lag of a velocity signal. Considering the periodic boundary conditions for all surfaces of the domain, we calculate the autocorrelation $R_{u_i u_i}^e (\Delta x_j)$ only for half of the computational domain. The vertical direction is parallel to the $x$ axis, while the horizontal component of the velocity is represented by the $y$ and $z$ components. To calculate $R_{ll}^e(\Delta l)$ we separately calculated $R_{vv}^e(\Delta y)$ and $R_{ww}^e(\Delta z)$ and averaged them as $R_{ll}^e = (R_{vv}^e+R_{ww}^e)/2$, where $\Delta l$ corresponds to the shift in horizontal component $y$ and $z$, respectively for $R_{vv}^e$ and $R_{ww}^e$. 
%\bv{BV: You should comment on and refer to your discussion in § \ref{sec:results}.3, where you discuss the velocity fluctuations being a function of your domain size}
In order to determine an appropriate domain size, it is necessary to ensure that the velocity autocorrelation function reaches a regime where it fluctuates minimally around zero. Ideally, the time when autocorrelation reaches zero value should also be independent of the domain size.
Figure \ref{fig:autocor_phi005} illustrates the autocorrelation of the vertical component of the velocity on the vertical axis $R_{uu}^e(\Delta x)$ and the horizontal component of the velocity as a function of the horizontal dimension $R_{ll}^e(\Delta l)$. As can be seen from the figure that $R_{uu}^e(x)$ approaches near-zero values without noticeable oscillations for domains $SD50$ and larger. However, the initial slope of $R_{uu}^e(\Delta x)$ for $SD50$ is steeper by comparison with simulations $SD65$ and $SD100$, which appeared to be almost identical to each other. The nearly identical slope of $R_{uu}^e(\Delta x)$ for $SD65$ and $SD100$ indicates that the spatial correlation of velocity fluctuations is effectively independent of the domain size. In contrast, the autocorrelation of horizontal velocity ($R_{ll}^e(\Delta l)$) decays to zero more rapidly but becomes negative for all domain sizes. The largest horizontal domain ($SD65$) exhibits the weakest negative excursion of $R_{ll}^e(\Delta l)$ and a slight upward trend for $l/D > 25$. This behaviour suggests that an even larger horizontal extent would be desirable; however, due to computational constraints, we consider domains with a vertical size of at least $65D$ and an even larger horizontal extent to be adequate.
As discussed in § \ref{sec:vel_fluct}, the particle velocity fluctuations increase with increasing domain size if tripple periodic boundary conditions are used. This behavior indicates a growth of correlated flow structures, which in turn suggests that the correlations are expected to persist over larger spatial separations and reach zero at larger lags for larger domains.
Given that $SD65$ already possesses the largest horizontal dimension among the tested cases, it was selected as the reference configuration for the main simulation campaign.

\bibliographystyle{jfm}
\bibliography{jfm}

\end{document}